\documentclass[11pt,a4paper]{article}
\usepackage{jheppub}
\usepackage{float}
\usepackage{mathrsfs,bm}
\makeatletter
\let\std@footnotetext\@footnotetext
\let\@footnotetext\std@footnotetext
\makeatother
\newcommand{\be}{\begin{equation}}
\newcommand{\ee}{\end{equation}}

\def\bse{\begin{subequations}}
\def\ese{\end{subequations}}

\newcommand{\el}[1]{\label{#1}}
\newcommand{\er}[1]{\eqref{#1}}

\newcommand{\ci}[1]{}

\newcommand{\lb}{\left(}
\newcommand{\rb}{\right)}
\newcommand{\lc}{\left.}
\newcommand{\rc}{\right.}
\newcommand{\lsb}{\left[}
\newcommand{\rsb}{\right]}

\newcommand{\nn}{\nonumber \\}
\newcommand{\p}{\partial}

\newcommand{\cd}{\nabla}

\newcommand{\eq}[1]{(\ref{#1})}

\newcommand{\ba}{\begin{eqnarray}}
\newcommand{\ea}{\end{eqnarray}}
\newcommand{\bal}{\begin{align}}
\newcommand{\eal}{\end{align}}
\newcommand{\bay}[1]{\left(\begin{array}{#1}}
\newcommand{\eay}{\end{array}\right)}
\newcommand{\eg}{\textit{e.g.} }
\newcommand{\ie}{\textrm{i.e.}, }

\newcommand{\at}[1]{{\Big|}_{#1}}
\newcommand{\zt}[1]{\textrm{#1}}

\def\rmd{{\rm d}}
\def\xa{{\alpha}}

\def\xd{{\delta}}
\def\xD{{\Delta}}
\def\xe{{\epsilon}}

\def\xg{{\gamma}}
\def\xG{{\Gamma}}

\def\xl{{\lambda}}

\def\xo{{\omega}}
\def\xO{{\Omega}}

\def\xs{{\sigma}}

\def\xt{{\theta}}

\def\xT{{\Theta}}

\def\CH{{\cal H}}

\def\CK{{\cal K}}
\def\CL{{\cal L}}
\def\CM{{\cal M}}

\def\CO{{\cal O}}

\title{Entanglement renormalization and integral geometry}

\author[a]{Xing Huang}
\emailAdd{xingavatar@gmail.com}
\affiliation[a]{Department of Physics, National Taiwan Normal University, Taipei, 116, Taiwan}
 \author[a]{and Feng-Li Lin} \emailAdd{linfengli@phy.ntnu.edu.tw} 
  
\abstract{We revisit the applications of integral geometry in AdS$_3$ and argue that the metric  of the kinematic space can be realized as the entanglement contour, which is defined as the additive entanglement density. From the renormalization of the entanglement contour, we can holographically understand the operations of  disentangler and isometry in multi-scale entanglement renormalization ansatz.  Furthermore, a renormalization group equation of the long-distance entanglement contour is then derived. We then generalize this integral geometric construction to higher dimensions and in particular demonstrate how it works in bulk space of homogeneity and isotropy.}  

\begin{document}

\maketitle

\section{Introduction}
As suggested by the AdS/CFT correspondence, spacetime could be emerging from the renormalization group (RG) flow of the boundary CFT. The radial direction in the bulk is the embodiment of RG flow. It remains unclear how geometry in the bulk can be reproduced from data on the boundary without using any pre-existing information provided by the bulk side, like a bulk-to-boundary propagator. Ever since the seminal work by the Ryu and Takayanagi \cite{Ryu:2006bv,Ryu:2006ef}, the relation between geometry in the bulk and the entanglement of the boundary state has been studied extensively (see \eg \cite{VanRaamsdonk:2009ar,VanRaamsdonk:2010,bianchimyers,Maldacena:2013xja} for general discussions relevant to this subject). However, a working prescription on building the bulk spacetime from boundary entanglement remains much desired.

    A promising framework to achieve the above goal is an efficient numerical algorithm proposed for evaluating the ground states of many-body systems by evolving the RG flow through the network of local unitary operations (LUs), i.e., the quantum circuit. This algorithm is called the multi-scale entanglement renormalization ansatz, denoted as MERA \cite{MERA1,MERA2} (and its continuous version cMERA \cite{Haegeman:2011uy}) and is a revised version of real space renormalization of the quantum state by also removing local entanglement (the LUs for such kind of functions are called disentanglers in MERA) besides coarse graining (the corresponding LUs are called isometries in MERA).  A keen observation in \cite{Swingle:2009bg} about the resemblance between the geometric structures of quantum circuit for MERA and the AdS space initiates further elaborated works 
\cite{Evenbly,Swingle:2012,Nozaki:2012zj,Mollabashi:2013lya,Qi:2013caa,Lee:2015vla,Pastawski:2015,Bao:2015uaa} along this direction.

  However, very little is known about the explicit form of the LUs in MERA except in simple free models \cite{Haegeman:2011uy,Nozaki:2012zj,Qi:2013caa,Lee:2015vla} based on variational principle.   Nevertheless, we expect the modification of the entanglement structure due to the LUs at every step of RG will be encoded in the increment of bulk geometry being generated along the radial direction. This leads to a recent proposal in \cite{Miyaji:2014mca,Miyaji:2015yva} (see also \cite{Miyaji:2015fia,Verlinde:2015qfa,ooguri}) that each (open/closed) bulk surface corresponds to a quantum (mixed/pure) state at a particular RG scale set by the radial position of the surface. Moreover, states at different RG scales are related by LUs of MERA along the RG trajectory. This new proposal takes into accounts the entanglement renormalization in contrast to the earlier works \cite{Hamilton:2006,bilson1,bilson2,Spillane,Bousso:2012,Czech:2012} on the holographic reconstruction of the bulk geometry.

    To refine the connection between the bulk geometry and boundary entanglement renormalization in more microscopic sense,  one needs some bulk geometric picture for the spatial distribution of entanglement so that the action of LUs of MERA on it generates the bulk line element along the RG trajectory. Besides, the entanglement entropy for some region can be understood as the additive sum of the contribution from the pairs of points with one end inside and one outside.  This provides a microscopic view point of entanglement entropy, a concept called entanglement contour as introduced in \cite{Chen:2014ec}. In the current description, the entanglement contour characterizes the fraction of each pair to the total entanglement entropy. 
       
  In this sense, the recent developments in differential entropy \cite{Balasubramanian:2013b,Balasubramanian:2013lsa,Myers:2014jia,Czech:2014wka,Headrick:2014eia,Czech:2014ppa} offers powerful tools ready for this purpose. Especially, a special way of discussing the distance measured in 3D hyperbolic space by the the integral geometry (see \eg\cite{santalo}) leads \cite{Czech:2015qta} to express the differential entropy (the length of the holographic target bulk curve) as the volume of the kinematic space of the geodesics intersecting the target bulk curve. Moreover, each point in the kinematic space by itself corresponds to a geodesic with two boundary end points. The end points can be intuitively and holographically associated with two infinitesimally small spatial regions on the boundary, and their entanglement contribution to the differential entropy is weighted by the volume measure of the kinematic space.   This picture will serve as the starting point of this work. 
  
     In this paper, we will adopt the framework introduced in \cite{Czech:2015qta} and take the tools therein to explore the local and microscopic understanding of the relation between bulk line element and the entanglement renormalization under the quantum state RG flow of the boundary CFT.  We divide the entanglement entropy of a point (or rather infinitesimal spatial region) with its complement into contributions from every other point and consider each piece as  a new type of additive measure of spatial entanglement. This quantity, labeled by a pair of points corresponds to the metric in the kinematic space (with the points serving as coordinates). It is very similar yet not exactly the same as the entanglement contour introduced in \cite{Chen:2014ec}. For clarity, we will anyway use the the term entanglement contour but the reader should keep in mind their minor discrepancy. The additive feature of the entanglement contour enables us to study its renormalization (holographically a point in the kinematic space of integral geometry) in a RG step (by advancing the boundary cutoff surface).
     
       In this microscopic picture we find that the holographic differential entropy at particular RG scale $z=\tilde z$ ($z$ is also the radial coordinate) measures the long-distance entanglement contour, i.e., contributed by the entanglement contour of pairs with end-points separated by distance larger than $2\tilde z$. As the area of the bulk surface for the differential entropy is unaffected by shifting the boundary, this implies that the associated long-distance entanglement contour is preserved under RG flow.  This is consistent with the picture of MERA in which disentangler only removes the short-distance entanglement. Based on this fact, we can derive the RG equation for the corresponding entanglement contour. 
        
   We furthermore generalize the above construction to (a spatial slice of) general $(d+1)$-dimensional static spacetime, for which a point in the kinematic space is a $(d-1)$-dimensional surface that consists of all the geodesics passing through it.  The volume form in the kinematic space can be determined by an invariant $(2d-2)$-form in the phase space of all geodesics and in the end-point coordinates, it is again given by the second derivatives of the geodesic length. The metric element can be reconstructed using the Crofton's formula in integral geometry and we explicitly demonstrate the construction using a homogeneous and isotropic space. Moreover, the correspondence between the kinematic volume element and the entanglement contour between two points can be generalized to higher dimensions straightforwardly. Hopefully this will offer a nice prescription on how to construct geometry from the entanglement structure.

The layout of the paper is as follows. In \S~\ref{sec:review}, we review the basics of integral geometry. In \S~\ref{sec:EnReno}, the connection between integral geometry and the renormalization group flow of entanglement contour is discussed. We explore the possible applications of integral geometry in higher dimensions in \S~\ref{sec:IGHigherD}. We conclude and discuss some of the future directions in \S~\ref{sec:conclusion}. Some of the relevant materials are relegated to the appendices.  

\section{Integral geometry in AdS/CFT}
\label{sec:review}
Integral geometry offers helpful tools in reconstructing the bulk geometry from boundary data in the context of AdS/CFT correspondence. In \S~\ref{subsec:review} we will first briefly review the basics of integral geometry and demonstrate how it works in pure AdS$_3$. We will follow, though not very closely the discussion in \cite{Czech:2015qta} (see also \cite{Lin:2014hva} for different applications of integral geometry). In \S~\ref{subsec:generic3d} we will then generalize the results obtained in \cite{Czech:2015qta} and compute the distance between two points in a generic AdS$_3$ space. In \S~\ref{subsec:metricfromE}, we consider the special case in which the two points are infinitesimally apart and their distance becomes the line element. We will further study the line element in later sections.

\subsection{Review of integral geometry}
\label{subsec:review}
In the framework of integral geometry, the volume of a $q$-dimensional object $M^q$ in a $d$-dimensional
symmetric space $\CM$ can be measured by $r$-dimensional planes. More precisely, the volume of $M^q$ can be expressed in terms of an integral over an auxiliary space $\CK$ called kinematic space, which consists of all the $r$-planes. In the context of AdS/CFT, we apply integral geometry to a $d$-dimensional spatial slice of the stationary bulk spacetime. In AdS$_3$ ($d=2$), the planes we need to consider are geodesics ($r=1$) and we are mainly interested in the bulk curves ($q=1$), whose lengths can be written in terms of integrals via the Crofton's formula ($G$ is the Newton constant of bulk gravity)
\be
\el{Crofton}
\frac{\xs(\xg)}{4G} = \frac 1 4 \int_{\xg \cap \xG \ne \emptyset} N (\xg \cap \xG )\; \xe_{\CK}\,,
\ee
where $N(\xg \cap \xG )$ denotes intersection number between the bulk curve $\xg$ and oriented geodesics $\xG$ ending on the boundary, and $\xe_\CK$ denotes the volume form of the kinematic space $\CK$.

In the case of pure AdS$_3$ the measure follows from that of the spacetime isometry group and has a natural entropic interpretation 
\be
\el{volumeform}
\xe_\CK(u,v) = \frac {\p^2 S(u,v)}{\p u \p v} \rmd u \wedge \rmd v\,.
\ee
The function $S(u,v)$ is the length (divided by $4G$) of the geodesic with $u,v$ being the boundary end
points, which are now also the coordinates of the kinematic space. According to Ryu-Takayanagi (RT) formulation \cite{Ryu:2006bv,Ryu:2006ef}, $S(u,v)$ is the holographic entanglement for the the boundary interval $[u,v]$, which is also denoted as $S_{EE}(u,v)$ (or simply $S_{EE}(|u-v|)$ for a translationally invariant state). The second derivative of $\p_u \p_v S$ can be understood as the limiting case of the conditional mutual information defined as
\be I(A,C|B):= S(AB)+S(BC)-S(ABC)-S(B)\,.\ee
In the current case, we choose
\[A=(u-du, u),\quad B=(u, v),\quad C=(v, v+dv)\,,\]
and hence $I(A,C|B) = \p_u \p_v S\; \rmd u \rmd v$,
which is positive definite as a result of strong subadditivity. 

The causal structure of kinematic space tells that two geodesics sharing the same left or right end point are null separated,\footnote{Following \cite{Czech:2015qta} a geodesic $\xg_1$ is in the future of $\xg_2$ if the interval of $\xg_1$ contains that of $\xg_2$.} and hence the end points can be used as null coordinates. As a result, we have the following metric in the kinematic space
\be
\el{metrickinematic}
\rmd s^2 = \frac {\p^2 S(u,v)}{\p u \p v} \rmd u \rmd v\,.
\ee
Substituting in the volume form \eq{volumeform}, the right hand side of \eq{Crofton} becomes
\be
\el{diffIG}
\int_{\xg \cap \xG \ne \emptyset} \frac {\p^2 S(u,v)}{\p u \p v} \rmd u \wedge \rmd v = -\oint \frac {\p S(u,v(u))}{\p u } \rmd u \,.
\ee
The integral is reduced to a surface integral over geodesics $v(u)$ tangent to the
bulk curve and this is essentially the differential entropy \eq{EAction}
(see Appendix~\ref{sec:dereivew} for a review on differential entropy). Differential entropy $E$ can be understood as the continuum limit ($n\to \infty$) of the following quantity
\be
\label{discretE}
E = \sum_{k=1}^n [S_{EE}(I_{k})- S_{EE}(I_{k+1} \cap I_{k})]\,,
\ee
where $I_k$ denotes the intervals that cover the boundary and $S_{EE}(I_k)$ is the entanglement entropy for each of them. We note that differential entropy is a quantity defined entirely from the data on the boundary.

We note however that at this point the set of geodesics intersecting $\xg$ cannot be determined without further input from the bulk. To see how the
geometry of the (bulk) real space is entirely encoded in the
kinematic space, the first step to take is to construct bulk points in the kinematic space. When a bulk curve shrinks to zero in size, it turns into a point and intuitively
we can identify the set of geodesics passing through it as the correspondence
in the kinematic space. This set for a point $A$ is a $(d-1)$-dimensional subspace
$p_A$ (called point curve) in the kinematic space.

To be concrete, we will write down the specific forms of geodesics and point
curves in pure AdS$_3$. In the global coordinate, the metric reads
\be
-\lb 1+\frac {R^2}{L^2}\rb \rmd T^2+\lb 1+\frac {R^2}{L^2}\rb^{-1} \rmd R^2+R^2 \rmd \xt^2\,.
\ee
A point in AdS$_3$ bulk can be mapped to a cover of the boundary $\{I[\xa(\xt)]\}$ (a collection of intervals $I[\xa(\xt)]$
whose union $\cup_\xt I[\xa(\xt)]$ is the boundary), where $\xt$ is the angular
coordinate of the center of an interval and $\xa(\xt)$ is half of its length. The function $\xa(\xt)$ contains all the necessary information to specify the intervals and hence we use it as the label. More specifically, a geodesic passing through $(R,\tilde \xt)$ subtends an interval centered at $\xt$ and the half length of this interval $\xa(\xt)$ is given by
\be
\el{alpha}
\alpha(\xt) =\cos^{-1}\left[\frac{R \cos (\theta -\tilde \theta)}{\sqrt{L^2+R^2}}\right]\,.
\ee 
The corresponding entanglement entropy of the interval is given by 
\be
\el{entropyAdS3}
S_{EE}(\xa) = \frac L {2 G} \log \frac {2L\sin \xa} {\mu} \,,
\ee
where $\mu$ is the UV cutoff. We can rewrite $S_{EE}$ in terms of end points $u,v$ and they are related to $\xa, \xt$ by
\be
\xt = \frac {u+v} 2,\quad \xa = \frac {v -u} 2 \,. 
\ee
Plugging \eq{entropyAdS3} into \eq{metrickinematic}, we get the metric in
the kinematic space 
\begin{equation}
\el{metricuv}
\rmd s^2 = \frac{L}{4G}\, \frac{\rmd u\, \rmd v}{2 \sin^2(v-u)/2}\,,
\end{equation}
The point $A$ is then described by a point curve $p_A$ given by $v(u)$ 
\be
\el{pointcurve}
v(u) = 2 \cos ^{-1}\left[\frac{\sqrt{L^2+R^2} \sin \left(\frac{u}{2}\right)+R \sin \left(\frac{u}{2}-\tilde \theta \right)}{\sqrt{L^2+2 R^2-2 R \sqrt{L^2+R^2} \cos (u-\tilde \theta )}}\right]\, .
\ee 
It appears that we again need the information from the bulk \eq{alpha}. However,
as demonstrated in \cite{Czech:2015qta}, the point curves can be constructed
iteratively. We will come back to this point later.

\begin{figure}[tbp]
\begin{minipage}[t]{0.5\textwidth}
\includegraphics[scale=0.5]{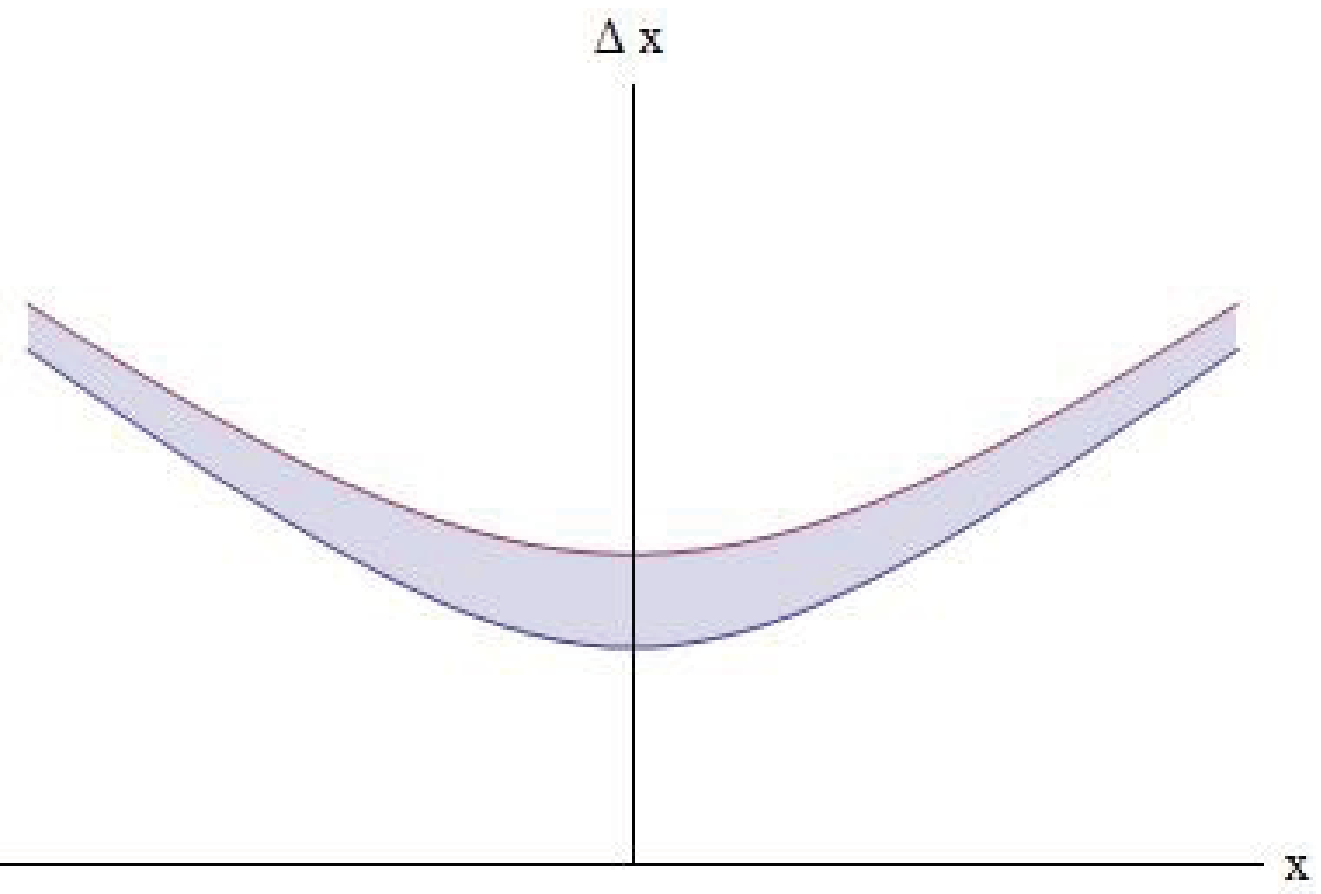}
\end{minipage}
\begin{minipage}[t]{0.5\textwidth}
\includegraphics[scale=0.5]{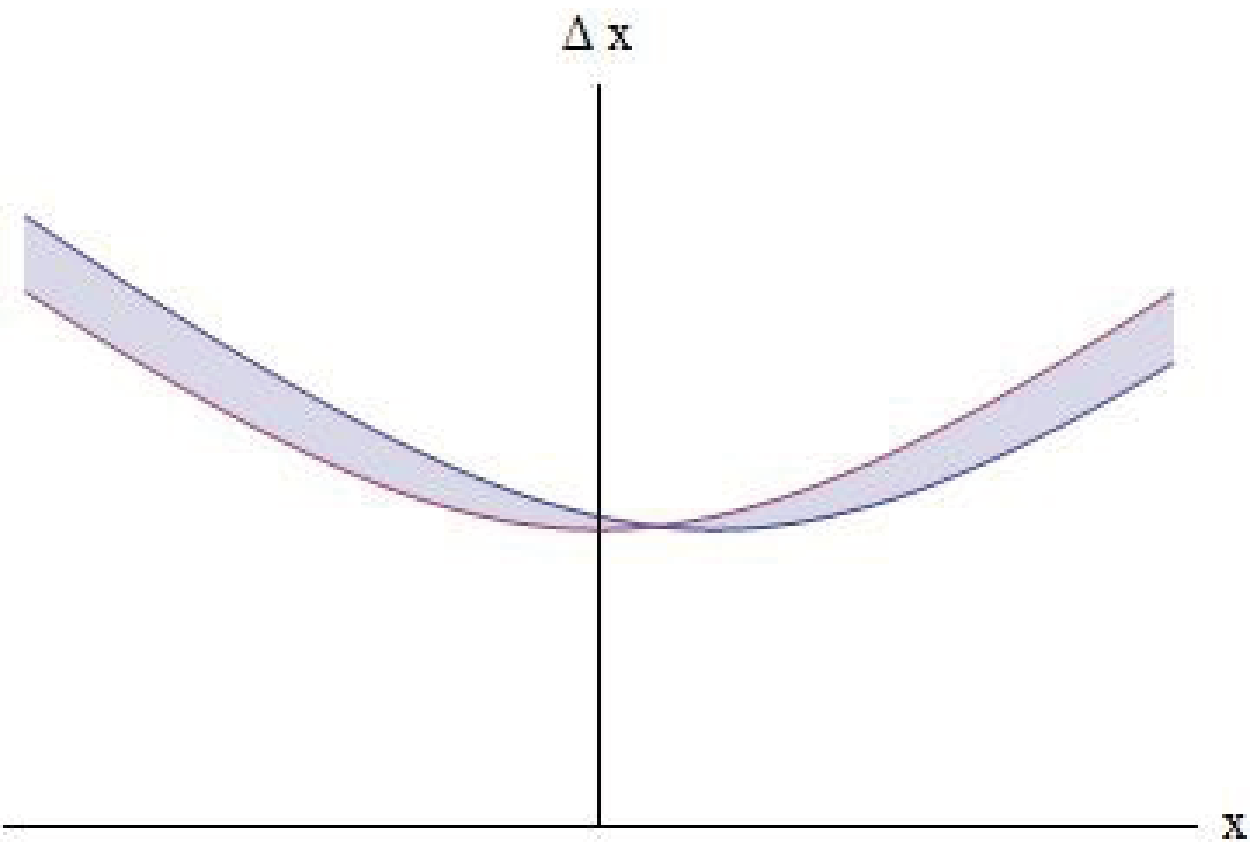}
\end{minipage}
\caption{Examples of point curves. Coordinates of $x$ (midpoint) and $\xD x$ (size) instead of end points $u,v$ are used to parameterize a geodesic. On the left $A=(\tilde x, \tilde z)$ and $B=(\tilde x, \tilde z+ d\tilde z)$ and on the right $A=(\tilde x, \tilde z)$ and $B=(\tilde x+ d\tilde x, \tilde z)$. In either case, $\tilde p_A\Delta \tilde p_B$ is indicated by the shaded region.}\label{fig5}
\end{figure}

We can define in $\CK$ a region $\tilde p_A$, called the future of $p_A$ which
consists of geodesics that will sweep through $A$ when being deformed to the boundary.
They fill up the region above the curve $p_A$. The distance between two points $A,B$ again follows from \eq{Crofton}. Only geodesics that intersect the geodesic between $A,B$ contribute to the length and they all lie in the region
sandwiched by the two curves $p_A, p_B$. More precisely they are specified by (see Fig.~\ref{fig5} for specific examples)
\be
\el{pointcurvedis}
\tilde p_A\Delta \tilde p_B \equiv (\tilde p_A\cup \tilde p_B) - (\tilde p_A\cap \tilde p_B)\,.
\ee
The distance then takes the following form
\be\label{integral-interval}
\frac {\rmd(A,B)} {4G}=  \frac 1 4 \int_{\tilde p_A\Delta \tilde p_B}\xe_\CK \,.
\ee
In higher dimensions, the Crofton's formula can be generalized \cite{santalo},
\be
\el{IntegralMeasure}
\int_{M^{q} \cap L_r \ne \emptyset} \xs_{q+r-d} (M^{q} \cap L_r )\; \xe_\CK  =\frac{O_{d}\dots
O_{d-r}O_{q+r-d}}{O_r\dots O_1 O_0 O_q} \xs_q (M^q)\,,
\ee
where $\xs_n$ denotes the volume (area/length) of an $n$-dimensional object and $O_{n}$ is the area of unit $n$-sphere
\be
O_{n} = \frac{2 \pi ^{\frac{n+1}{2}}}{\Gamma \left(\frac{n+1}{2}\right)}\,.
\ee
Here we again use $\xe_\CK$ to denote the measure of the kinematic space $\CK$, which now consists of all the $r$-dimensional planes \footnote{The planes in \eq{IntegralMeasure} are unoriented. But for convenience, we often consider
oriented geodesic and put in a compensated factor like $1/2$ in \eq{Crofton} (the other 1/2 follows from $O_2/O_1$).} and the integration is over all the planes that intersect the $q$-dimensional surface $M^q$. Intuitively, each $r$-plane performs a measurement on the volume and sees the size of the cross section $\xs_{q+r-d} (M^{q} \cap L_r )$. The formula \eq{IntegralMeasure} then tells us that the total volume can be obtained by adding up all the pieces. Notice that we no longer introduce the factor $(4G)^{-1}$ in $\xe_\CK$ because generically $\xs_q (M^q)$ is not related to entanglement entropy. We will discuss integral geometry in higher dimensions in \S~\ref{sec:IGHigherD}. 

\subsection{Generalization to generic AdS$_3$ spaces}
\label{subsec:generic3d}
Although in \cite{Czech:2014ppa,Czech:2015qta} the Crofton's formula \eq{Crofton} was only fully tested in pure AdS$_3$ and its quotient (BTZ black holes), it is not difficult to see that the formula is valid in more general situations. Here we would like to use it to compute the distance between two points in a generic AdS$_3$ space. In this space the geodesics can be parameterized using their end points $u,v$ and the volume form is given by \eq{volumeform}. Points (or rather point curves) are given by one dimensional surfaces in the kinematic space. We consider two bulk points $A$ and $B$, and their point curves are denoted by $p_A$ and  $p_B$ respectively. We work in a Poincare patch and without lost of generality we also assume $v_A(u) < v_B(u)$ as $u \to -\infty$ (see \eg the right panel of Fig.~\ref{fig5} for an example of this kind). The distance between $A,B$ can be computed by summing over all the geodesics intersecting the geodesic which connects $A$ and $B$, i.e., by \eq{Crofton},
\ba
& &  \frac{\rmd(A,B)} {4G}  = \frac 1 4 \int_{-\infty}^{+\infty} \left| \frac {\p S(u,v_B(u))}{\p u} - \frac {\p S(u,v_A(u))}{\p u}\right| du \nn
& = & \frac 1 2 \lb \int^{u(p_A \cap p_B)}_{-\infty}-\int_{u(p_A \cap p_B)}^{+\infty}\rb \lsb \frac {\p S(u,v_B(u))}{\p u} - \frac {\p S(u,v_A(u))}{\p u}\rsb du\nn
& := & \lsb S_B(u)-S_A(u)\rsb \at{u(p_A \cap p_B)}\,,\label{line-element-generic}
\ea
where $p_A \cap p_B$ is the geodesics passing through both points $A$ and $B$, i.e., corresponds in kinematic space to the intersection point of the right panel of Fig.~\ref{fig5}, and $u(p_A \cap p_B)$ is its left end point on the boundary. In the last step, we define $S_{A,B}(u)$ as the length (divided by $4G$) of the geodesic from the bulk
point $A$ (or $B$) to $u$ and use the following relation
\be
\frac{\p S(u,v_{A(B)}(u))}{\p u} = \frac {d S_{A(B)}(u)}{d u}\,,
\ee
which is continuous in either branch. We note that in contrast to \eq{Diffbound} the derivative with respect to one of the end points becomes a total derivative when the other end point is held fixed. Note that $S_B(u)-S_A(u)$ is the difference in length of the two geodesics that start from $A$ and $B$ respectively but both end at $u$ on the boundary. Obviously, as the final expression of  \eq{line-element-generic} is evaluated on the geodesic $p_A \cap p_B$,  this (infinitesimal) difference is by construction the (line element) distance between $A$ and $B$. We note that the derivation above is independent of specific geometry of the (bulk) real space.

\subsection{Line element from integral geometry} 
\label{subsec:metricfromE}
 Previously we have discussed the construction of the length of a bulk curve from the integral geometry. A particular case for this construction is to obtain the line element, i.e., the metric of the infinitesimal interval, which is the building block of the bulk differential geometry. This is essentially the infinitesimal version of \eq{integral-interval}. For simplicity we will work in the Poincare coordinate so that the metric takes the form
\be\label{poincare-ads3}
ds^2={L^2 \over z^2}(dz^2+dx^2-dt^2)\;. 
\ee
We now like to obtain the line element between two infinitesimally separated bulk points along radial direction, i.e., $(\tilde{x},\tilde{z})$ and $(\tilde{x},\tilde{z}+\rmd z)$. By the integral geometry, we need the set of geodesics corresponding to the point curve for the bulk point $(\tilde{x},\tilde{z})$. These geodesics can be parametrized by the coordinate of the highest point $x$, for which the geodesic subtends an interval on the boundary of length 
\be
\el{coverPoincare}
\xD x =2 \tilde z \sqrt{\left(\frac{x-\tilde x}{\tilde z}\right)^2+1}\,.
\ee

   As the metric \eq{poincare-ads3} is translational invariant along the $x$-direction, it is more convenient to parametrized the kinematic space of the geodesics by $x$ and $\xD x$ instead of the null-coordinates $u,v$, i.e., $x=\frac 1 2 (u+v), \xD x=u-v$. Then, the measure factor $\partial_u\partial_v S$ becomes $-\p^2_{\xD x} S$ so that we obtain the line element from \eq{integral-interval} as following (see Fig.~\ref{fig5} for the region to be integrated):
\be \el{metricPoin}
\sqrt{g_{zz}}\rmd z= {4G}\, \rmd z \int_{-\infty}^{+\infty} \frac 1 2 \lb -\frac {d^2 S} {d\xD x^2}\rb \left|\frac {d\xD x} {d\tilde z} \right|\rmd x = \rmd z\int_{\tilde x}^{+\infty} \frac{L \tilde z\, \rmd x}{\left[(x-\tilde x)^2+\tilde z^2\right]^{3/2}} = \frac{L}{\tilde z}\rmd z\,.
\ee
Here we introduce the absolute value for the difference in $\xD x$ between two points to account for the positive volume.  Similarly, the generic line element can be constructed as following: 
\be
\rmd s= -{4G}\rmd  z \int_{-\infty}^{+\infty} \frac 1 2 \frac {d^2 S} {d\xD x^2} \left|\frac {d\xD
x} {d\tilde{z}} \right|\rmd x  =\frac {\rmd z} 2 \int_{-\infty}^{+\infty} \frac{L |k(x-\tilde x)+\tilde z| \rmd x }{\left[(x-\tilde x)^2+\tilde z^2\right]^{3/2}} = \frac{\sqrt{1+k^2} L}{\tilde z} \rmd z\,.
\ee 
where $k := \p_{\tilde z} \tilde x$. 

\subsection{Line element from entanglement renormalization via AdS/MERA}
   The motivation of this work starts from the observation that the bulk line element constructed above via integral geometry can also be obtained from the holographic entanglement renormalization. This then suggests that entanglement renormalization  could also be understood from the integral geometry point of view.
   
   The idea of entanglement renormalization was first proposed in \cite{MERA1,MERA2} in the context of solving many-body systems by adopting a network of local unitary transformations (LUs). This algorithm is called multi-scale entanglement renormalization ansatz (MERA). Later on it was realized \cite{Swingle:2009bg,Evenbly,Swingle:2012} that the geometric structure of MERA network of gapless system is of AdS space. This equivalence is then coined as AdS/MERA correspondence, see also \cite{Nozaki:2012zj,Qi:2013caa,Lee:2015vla,Mollabashi:2013lya,Pastawski:2015,Bao:2015uaa} for later development.  

    Motivated by AdS/MERA, a more refined picture in the context AdS/CFT proposed recently \cite{Miyaji:2014mca,Miyaji:2015yva,Miyaji:2015fia} (see also \cite{Verlinde:2015qfa}) is the so-called surface/state correspondence (SS-duality), which we will adopt for our discussions in this work.  The proposal states that any co-dimensional two convex surface $\Sigma$ is dual to a holographic quantum state described by density matrix $\rho({\Sigma})$. Especially, if $\Sigma$ is closed, it is dual to a pure state $\rho({\Sigma})=|\psi(\Sigma)\rangle\langle \psi(\Sigma)|$.  In the context of AdS/CFT, the bulk space can be foliated into congruence of these convex surfaces, and different surfaces/states are connected by a network of LUs which can be interpreted as renormalization flow if one follows the UV to IR directions. The LUs of MERA are utilized to disentangle the IR-irrelevant degrees of freedom (d.o.f.) from the the IR-relevant ones.  By keeping only the IR-relevant d.o.f.,  the RG flow can be treated as a unitary transformation. The feature of disentangling the IR-relevant and irrelevant d.o.f. makes the RG flow as a process of entanglement renormalization. 
  
  Based on SS-duality,  We now demonstrate that the line element of AdS$_3$ can also be understood as the holographic entanglement renormalization, i.e., disentangler and isometry (both are LUs) in the MERA construction if one follows the AdS/MERA proposal.   The simplest way to see this is to consider the entanglement between a semi-infinite line interval  $[\tilde x, \infty)$ and its complement in holographic CFT$_2$. The holographic entanglement entropy is given by the length of the a vertical bulk line $x=\tilde{x}$, i.e., $\int \sqrt{g_{zz}} \rmd z$. If we move from a surface/state at $z$ to the one at $z+dz$, the length is then deduced by the amount $\sqrt{g_{zz}}\rmd z$ so that the entanglement across the point at $\tilde x$ decreases by the amount of ${1\over 4G} \sqrt{g_{zz}}\rmd z$.

    Next we consider a more complicated case for the entanglement removal of a line interval of length $\ell$. The holographic entanglement entropy for the surface/state $\psi(\tilde z)$ at scale $z=\tilde{z}$ can be obtained by the RT formula and the result is \footnote{Notice that the RG running of a state, or rather in this particular case the running of $S_{EE}(\ell;\tilde z)$ must agree with the geometry of the bulk as indicated by surface/state correspondence.}
\be
\el{EEz}
S_{EE}(\ell;\tilde z) = \frac L {2 G} \log \frac {{\ell \over 2}+\sqrt{ ({\ell \over 2})^2
+{\tilde z}^2}} {\tilde z}\,.
\ee    
Thus, the amount of the entanglement removal when we move the surface/state at $z=\tilde z$ to the one at $z=\tilde z +\rmd z$ is  given by
\be
\el{dSdz}
- \frac {dS_{EE}}{dz}\at{\tilde z} \rmd z = \frac L {4 G} \frac{\ell}{{\tilde z} \sqrt{({\ell \over 2})^2+{\tilde z}^2}} \rmd z\;.
\ee    
Interestingly, if we take the limit $\ell \to \infty$, then  \eq{dSdz} becomes
\be 
\el{metricfromEE}
- \frac {dS_{EE}}{dz}\at{\tilde z} \rmd z \to  \frac L {4 G}{2 \rmd z \over \tilde z} \,.
\ee 
The additional factor of two is due to the double counting of two sides of the interval. After deducting this factor we reproduce the line element as obtained from the integral geometry.  This agrees with the implication that the entanglement entropy being removed by LUs is converted into the line element. 

We can also apply integral geometry to the space enclosed by a closed convex surface.
Notice that unlike in \cite{Czech:2015qta}, for a general surface we need to parameterize the geodesics by their end points on the surface. For example,
if the surface is the slice $z=\tilde z$ then we use the corresponding $S(u, v;\tilde z)$ (which obviously is different from $S(u, v;0)$) to define the measure. It is non-trivial that the volume in the kinematic space can be expressed using the boundary data on this slice. We will prove this in \S~\ref{subsection:inv-K}.  The ability to define entanglement contour on different slices yet still identify it with the kinematic measure makes it possible to study the RG running of entanglement contour, which we will do later in this section.

\section{Refined entanglement renormalization and integral geometry in AdS$_3$}
\label{sec:EnReno}
 
    Motivated by the different constructions of the bulk line element  either from integral geometry method or from entanglement renormalization, in this section we will explore the connection between integral geometry and entanglement renormalization in AdS$_3$. We will then generalize the consideration to higher dimensional AdS space in the next section.  To proceed, we will first reconsider the Ryu-Takayanagi (RT) formula of the holographic entanglement form the integral geometry point of view. By this way, we recognize the kinematic measure can be understood as entanglement contour, i.e., an additive measure of entanglement density proposed in \cite{Chen:2014ec}. Furthermore, by adopting the SS-duality, we can see geometrically how the long-distance entanglement contour renormalizes under the change of RG scale. Especially, a RG equation of the long-distance entanglement contour will be proposed based on the fact that the local unitary operation of MERA can only remove short-range entanglement but keep the long-distance one intact.\footnote{Here the notions of long and short are only defined relatively compared to the RG scale. In a discrete picture, the short-distance entanglements refer to the those between neighboring sites while all others are viewed as long.}

\subsection{Entanglement contour associated with pair of points}
   
\begin{figure}[tbp]
\centering
\includegraphics[height=55mm]{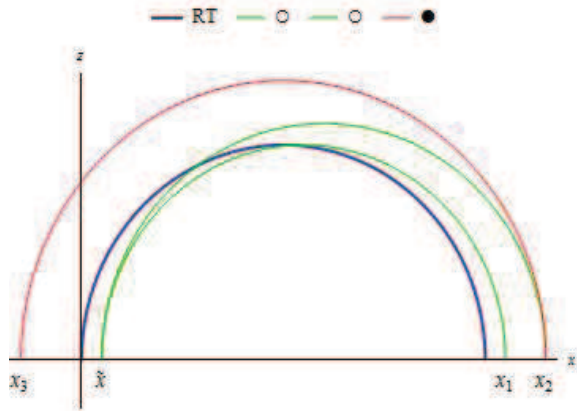}
\caption{We can visualize the kinematic space metric $\p_u \p_v S(u,v)$ as counting the amount of entanglement between pairs of points crossing the entangling surface, which contribute the entanglement entropy in an additive sense. Only the geodesics such as the green ones (starting one end at $\tilde x$) intersecting the RT geodesic (the blue one) contribute to the entanglement entropy of interval covered by RT geodesic, while the red one does not.}\label{figBell}
\end{figure}

   According to the RT formula, entanglement entropy in a two dimensional CFT can be computed from geodesic in the bulk. The latter is the special case of bulk curve and its area can be reproduced using the Crofton's formula \eq{Crofton}. More explicitly, we consider the surface state which is a horizontal surface at $z=\tilde z$ and the holographic entanglement entropy for the interval $I_{\ell}:=[\; \tilde x, \tilde x +\ell \;]$ can be rewritten as the sum of contribution associated with spatially separated pairs of points
\be
\el{Bellpair}
S_{EE}(\ell;\tilde z)= \frac 1 2 \int_{-\infty}^{\tilde x} \rmd u \int_{\tilde x}^{\tilde x+l} \rmd v \frac {\p^2 S(u,v;\tilde z)}{\p u \p v} + \frac 1 2  \int^{\infty}_{\tilde x+\ell} \rmd v \int_{\tilde
x}^{\tilde x+l} \rmd u \frac {\p^2 S(u,v;\tilde z)}{\p u \p v} \;.
\ee 
In the above either $u$ or $v$ is inside $I_{\ell}$ but not both.  The expression \eq{Bellpair} can be easily understood holographically as shown in  Fig.~\ref{figBell}: when using the integral geometry to obtain the length of RT geodesic $\Gamma_{\ell; \tilde z}$ for the holographic entanglement entropy, only the set of geodesics intersecting this RT one are counted. As $\Gamma_{\ell; \tilde z}$ has end points at $u=\tilde x$ and $v=\tilde x+\ell$ only the geodesics with one end-point in $I_{\ell}$ and another one outside will intersect $\Gamma_{\ell; \tilde z}$. \footnote{Due to the additive feature of \eq{Bellpair}, it is tempting to visualize the pairs of points as Bell pairs so that the entanglement entropy is counting the number of Bell pairs crossing the entangling surface. However, this visualization could be too naive to be compatible with the interpretation of the metric $\frac {\p^2 S(u,v;\tilde z)}{\p u \p v}$ the conditional mutual information.}

   In the limiting case $\ell\rightarrow dx$,  \eq{Bellpair} is now counting the entanglement of a point (with its complement) and can be reduced to
\be
\el{contourpoint}
S_{EE} (\rmd x;\tilde z) := \int_0^\infty \bar{s}(r,\tilde x;\tilde z) \rmd
r =\frac 1 2 \rmd x \lb\int_{-\infty}^{\tilde x} \frac {\p^2 S(u,v;\tilde z)}{\p u \p v}\at{v=\tilde x} \rmd u+ \int_{\tilde x}^{\infty}  \frac {\p^2 S(u,v;\tilde z)}{\p u \p v}\at{u=\tilde x} \rmd v\rb   \,,
\ee
Notice that the validity of this equation is independent of the Crofton's formula and follows from the fact that $S(u,v;\tilde z)$ is by definition entanglement entropy of an interval.

The integral \eq{contourpoint} shows the additive feature of the entanglement entropy in terms of the differential sum of the density quantity $\bar{s}(r,\tilde x;\tilde z)$, which measures the infinitesimal contribution to entanglement entropy parameterized by a pair of points located at $(\tilde{x},\tilde z)$ and $(\tilde{x}\pm r,\tilde z)$ with $r:=|u-v|$.   This kind of the quantity is recently proposed in \cite{Chen:2014ec} and is called ``entanglement contour" which should be positive and summed to the entanglement entropy (see also \cite{Nozaki:2013wia,Bhattacharya:2014vja} for a similar concept). Formally, entanglement contour $s_A (i)\ge 0$ is defined as contribution to the entanglement entropy $S_A$ from a subset $i \subset A$ ($\sum_i s_A(i) = S_A$). Instead,  $\bar{s}(r,\tilde x;\tilde z)$ of \eq{contourpoint} is defined for $i\subset \bar{A}$, here $A$ is the line element $\rmd x$ at $\tilde x$ and $\bar{A}$ is its compliment. However, we will abuse the terminology a bit and also call $\bar s$ the entanglement contour. As $S_{\bar A} = S_A$ for a pure state, in this case $\bar s$ is then the ordinary entanglement contour for the region $\bar A$.  \footnote{As we shall see later in \S~\ref{sec:IGHigherD}, the definition of entanglement contour and its correspondence with the kinematic measure can be carried over to higher dimensions even though the entanglement contour can no longer be understood as conditional mutual information.} An interesting implication of this microscopic picture of entanglement entropy is that the entanglement contour is definitely not bi-partite as the target point is correlated with all other points. \footnote{Some more minor constraints for the entanglement contour can be found in \cite{Chen:2014ec}. In particular, for a subset $X \subset A$, unitary transformation restricted to $X$ changes neither $s_A(X)$ or $s_A(A-X)$. This implies that $\p_u\p_v S(u,v;\tilde z)$ as entanglement contour is a quantity tied to a pair of points since it is invariant under the transformation on the region away from $u,v$. One can see this by noting that $S$ can also be computed from the two-point correlator of heavy operator. However, such property is not obvious if we instead view it as the conditional mutual information since the tripartite entanglement between $A,B,C$ in general is disrupted by a transformation involving $B$ and $\overline {ABC}$.}

   Based on \eq{contourpoint} we can define the long-distance entanglement contour (of a point) for latter usage, which only includes the contour of pairs separated by distance larger than some length scale $\ell_{\geq}$. That is, we just set the lower limit of the integral \eq{contourpoint} to be $\ell_{\geq}$, and define
\be
\el{thetadefinition}
\xT(\ell_{\geq},\tilde x; \tilde z)\rmd x:=\int_{\ell_{\geq}}^\infty \bar{s}(r,\tilde x;\tilde z) \rmd r = \frac 1 2 \rmd x \lb \frac {\p S(\tilde x-\ell_{\geq},v;\tilde z)}{\p v}\at{v=\tilde x}  - \frac {\p S(u,\tilde x+\ell_{\geq};\tilde z)}{\p u}\at{u=\tilde x} \rb \,.
\ee
We note that this quantity by definition is another entanglement contour since $\bar s(i\cup j) = \bar s(i)+\bar s(j)$.  The first term of the entanglement contour $\bar{s}(r,\tilde x;\tilde z)$ (contribution from the segment $\rmd x$ at $\tilde x$) agrees with the entanglement contour of the semi-infinite interval $I_{\ell \to \infty}$ as defined below. 

\begin{figure}[tbp]
\centering
\includegraphics[height=45mm]{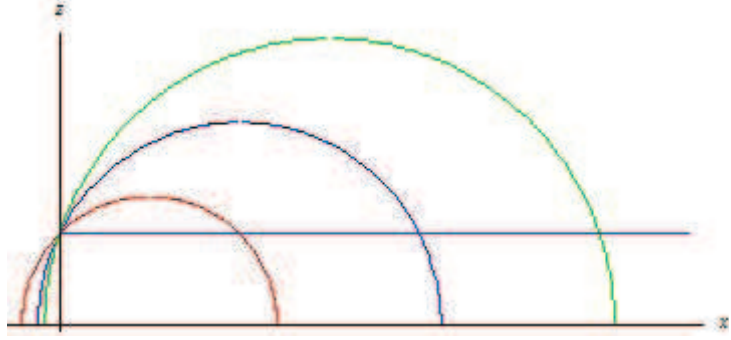}
\caption{Various geodesics passing through the point at $z = \tilde z$. When the boundary is lifted to $z=\tilde z$, they become curves sharing the same end points and cover the successive intervals denoted as $\{ I_k\}$, which can be used to define differential entropy \eq{discretE}. In this case, the longer curves (\eg the green one) always contain the shorter (the red and blue ones) entirely (\ie $I_1\subset I_2\subset \cdots I_k\subset I_{k+1}\subset \cdots \subset I_{\ell}$  so that  $S_{EE}(I_k)-S_{EE}(I_{k+1}\cap I_k)=0$). From the perspective of the boundary (now at $z=\tilde z$), this implies the differential entropy is given by the length of the longest geodesic. }\label{fig0}
\end{figure}

 The holographic entanglement contour for the entanglement entropy of an interval $I_{\ell}$ of length $\ell$ for a surface/state $\psi(z)$ (at scale $z$) can be constructed similarly in the way of integral geometry, i.e., using \eq{Bellpair}. In the limit of $\ell \to \infty$, we have 
\be\label{SeeDiff}
S_{EE}(x > \tilde x ;\tilde z):= \frac 1 2 \lim_{\ell \to \infty} S_{EE}(\ell;\tilde z) =\int_{\tilde x}^{\infty} s(\tilde x, x;\tilde z) \rmd x = \frac 1 2 \int_{\tilde x}^{\infty} \partial_{x} S(\tilde x, x;\tilde z) \rmd x\;.
\ee
This integral is performed over half of the point curve of a point at $(\tilde x, \tilde z)$ (see Fig.~\ref{fig0}) and gives half of the differential entropy for the geodesic $\lim_{\ell \to \infty} \Gamma_{\ell; \tilde z}$ \cite{Czech:2014tva}. Here the entanglement contour $s(\tilde x, x;\tilde z)$ follows the usual definition, i.e.,  $\sum_{i\in A} s_A(i) =S_A$.

  Note that, rewriting the entanglement entropy in terms of differential entropy also provides us a way to understand \eq{metricfromEE} (\ie renormalization of entanglement entropy matches the length of line element) from the viewpoint of integral geometry, i.e., half of the difference between the differential entropies of the two point curves gives their distance. Notice that we need to switch to the midpoint coordinate in order to rewrite \eq{SeeDiff} as differential entropy of a point, producing an extra factor of $2$.

  At this moment, it is interesting to compare the entanglement contour holographically obtained here (in the UV limit) with the one for the gapless system on the 1D lattice solved numerically in \cite{Chen:2014ec}.  For pure AdS$_3$, the holographic entanglement contour for $I_{\ell\rightarrow \infty}$ is
\be
\el{contourads3}
s(|x-\tilde x|;\tilde z):=s(\tilde x, x;\tilde z)={L\over 8G} {1\over \sqrt{(x-\tilde x)^2+{\tilde z}^2}}\;,
\ee
where $\tilde z$ is regarded as the UV cutoff. In the UV limit $\tilde z \rightarrow 0$, the entanglement contour $s(|x-\tilde x|;\tilde z=0) \propto 1/|x-\tilde x|$, this agrees with the result in \cite{Chen:2014ec}. Moreover, it is easy to see the aforementioned fact that $\Theta(\ell_{\geq},\tilde x;\tilde z)=2 s(|x-\tilde x|:=\ell_{\geq};\tilde z)$. 

  In this subsection, we introduce the notion of the entanglement entropy of a point with its complement . This notion seems peculiar but provides a very refined picture. In the next subsection we see that its peculiarity helps to consider the entanglement renormalization in the context of SS-duality. Especially, we can derive a holographic RG equation for $\xT(\ell_{\geq},\tilde x; \tilde z)$.

\subsection{RG equation of long-distance entanglement contour}
\label{subsec:econtour}

    In the previous discussions we have elaborated the relation between the entanglement entropy and the bulk geometry from the integral geometry approach. Especially, bulk line element can be reproduced from the infinitesimal RG flow of the entanglement entropy, and the vice versa the entanglement contour can be arrived from the bulk integral geometry for which a new microscopic measure of entanglement is deduced. It would be nice if further information about the surface/state $\psi(z)$ of SS-duality as defined in \cite{Miyaji:2015yva,Verlinde:2015qfa,Miyaji:2015fia} can be obtained from the entanglement contour, of which the RG flow should be related to the LUs of the corresponding MERA network.  Unfortunately, a definite recipe of constructing these LUs based on entropic consideration remains out of reach. Despite that, it does not prevent one from studying the entanglement renormalization in the context of AdS/MERA duality. 

 One interesting question we would like to answer is how the LUs under the RG flow affect the entanglement contour.  Intuitively, the LUs only act on the contour of short-ranged pairs so that these operations under RG flow can only remove the short-distance entanglement but keep the long-distance one intact. This is consistent with the fact that $\xD x$ in \eq{coverPoincare} does have a minimum $2\tilde z$ for a surface state at $z=\tilde z$. However, a typical RG running behavior of the entanglement contour given in \eq{contourads3} seems to contradict to the local nature of entanglement renormalization as (i) the entanglement contour is finite even for $x-\tilde x<\tilde z$ so that the short-distance entanglement is not removed completely; and (ii) the entanglement contours of all length scales decrease in the RG flow so that the long-distance contour seems to not remain intact under the local unitary operations.  We will see that these contradictions are just illusion caused by introducing the cutoff in a continuous theory. 
 
  Holographically, it is easy to understand the entanglement contour for the long-distance entanglement, i.e., long-distance entanglement contour. Let us consider the differential entropy associated with the bulk line $z=\tilde z$. According to the integral geometry representation of the differential/entanglement entropy such as the one given in \eq{contourpoint}, only the contour associated with pairs separated by distance greater than $2\tilde z$ will have nonzero contribution as the geodesic connecting the pair can intersect the $z=\tilde z$ line. Formally, the corresponding long-distance entanglement contour for the surface state $\psi(z=0)$ is given by the quantity $\xT(2\tilde z,\tilde x; 0)$ previously defined in \eq{thetadefinition}.  
   
\subsubsection{Field theory point of view}   

   Before we study the holographic RG flow of long-distance entanglement contour in full details, we would like to give a heuristic picture of the long-distance entanglement contour for a surface/state at RG scale $z=\tilde z$ from the field theory point of view. Here the corresponding entanglement contour contributing to the differential entropy is labelled by  a pair of points $a$, $b$ separated by a distance $\ell_{\geq}$. Despite of using the bi-partite labelling, the entanglement contour by its nature is not bi-partite as emphasized before.

\begin{figure}[tbp]
\begin{minipage}[t]{0.5\textwidth}
\includegraphics[height=25mm]{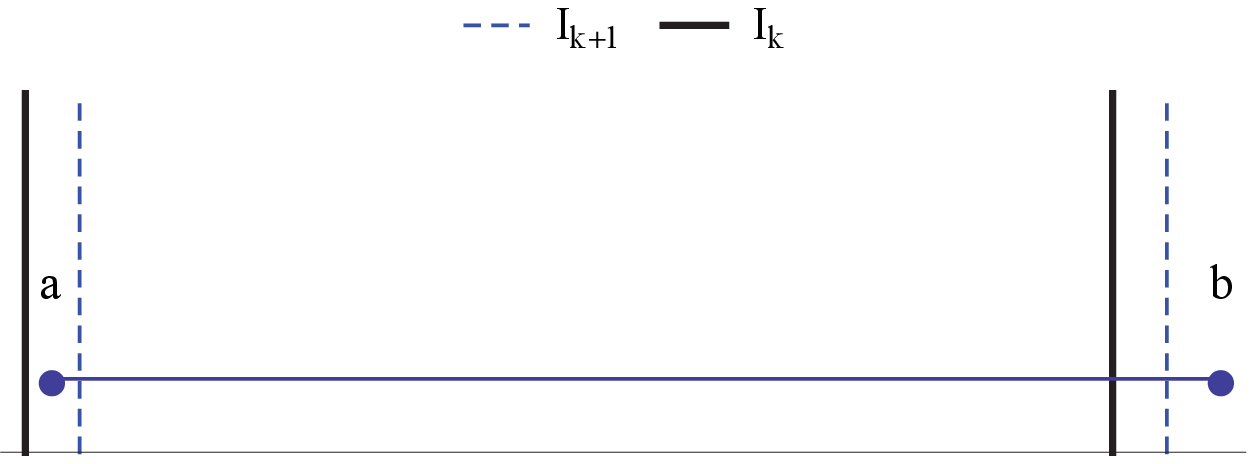}
\end{minipage}
\begin{minipage}[t]{0.5\textwidth}
\includegraphics[height=25mm]{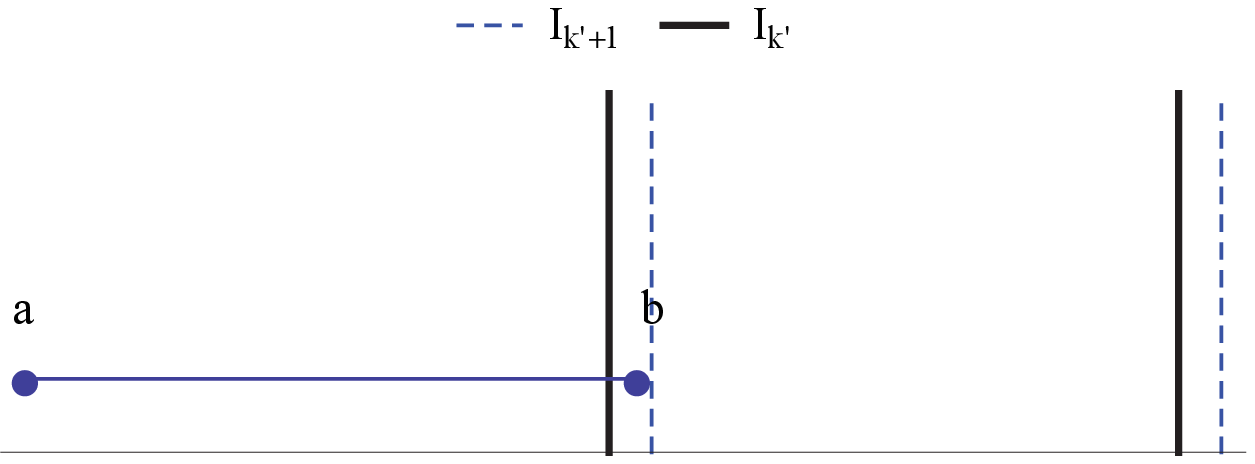}
\end{minipage}
\caption{ For a pair of points $a$ and $b$ separated by a distance larger than twice of the RG scale $\tilde z$, there are two dual situations of partition intervals for some $k$ and $k'$ shown here. The combined contributions to the different entropy form these two dual situations is positive.  See main text for detailed description. Note that for better vision we move the points $a,b$ off the axis.}\label{fig1}
\end{figure}

   The set of intervals for evaluating the differential entropy is denoted by $\{ I_k \}$ as usual. For the current case, the length of each $I_k$ is $2\tilde z$, i.e., for a surface/state at RG scale $z=\tilde z$.
For $\ell_{\geq}> 2 \tilde z$, then there should exist two corresponding situations as depicted in Fig. \ref{fig1}. The left panel is the situation that it happens to have a particular interval $I_k$  so that 
 $a \in I_k - I_{k+1}$ and $b \notin I_k$. Here we assume $a$ is to the left of $b$ and the center of $I_k$ is also to the left of the center of $I_{k+1}$).  According to \eq{contourpoint}, the contour of this pair contributes to $S_{EE}(I_k)$ but not to $S_{EE}(I_k \cap I_{k+1})$, that is, it has positive contribution to  $S_{EE}(I_k)$ - $S_{EE}(I_k \cap I_{k+1})$, i.e., the $k$-th fractional part of the differential entropy. At the same time, there should also be a particular $I_{k'}$ for which the same pair is in the situation as depicted in the right panel of Fig. \ref{fig1}: $b \in I_{k'} - I_{k'+1}$ and $a \notin  I_{k'}$ \footnote{To be more precise, before taking the continuous limit pairs of size $\ell_{\geq}>2\tilde z-\xD_{I_k - I_{k+1}}$ can contribute but in the limit $\xD_{I_k - I_{k+1}} \to 0$.} so that the pair also has positive contribution to the differential entropy. 

\begin{figure}[tbp]
\begin{minipage}[t]{0.5\textwidth}
\includegraphics[height=25mm]{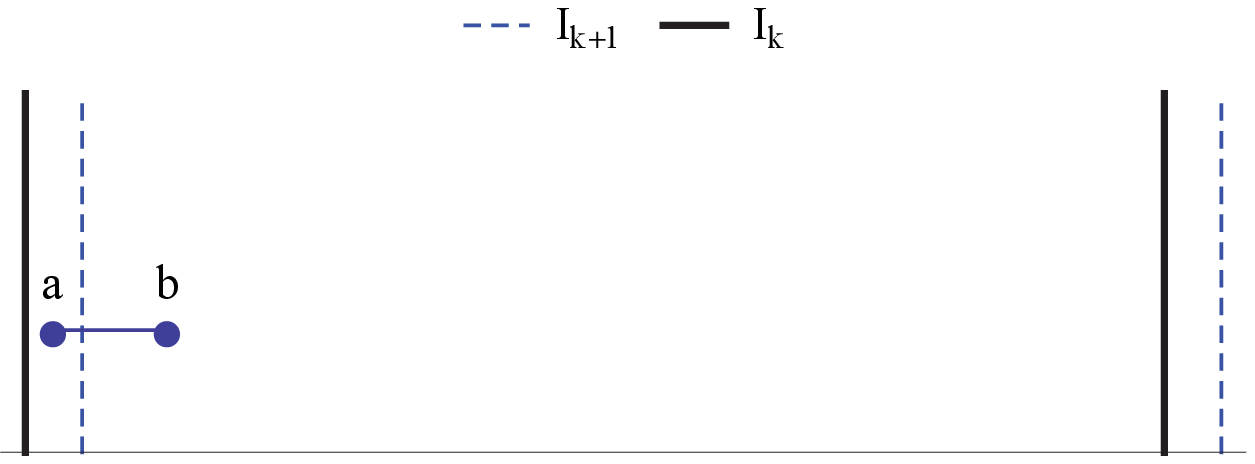}
\end{minipage}
\begin{minipage}[t]{0.5\textwidth}
\includegraphics[height=25mm]{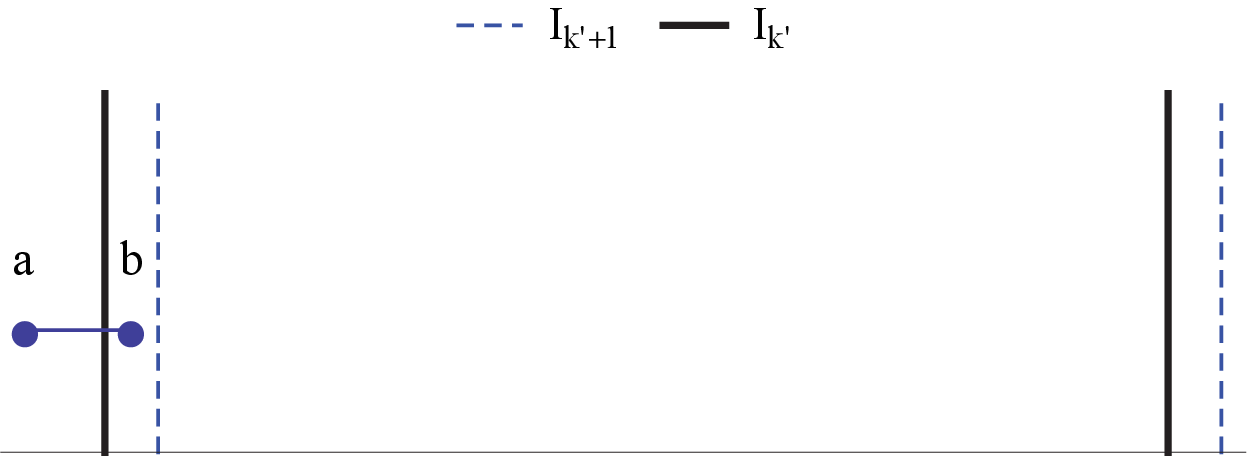}
\end{minipage}
\caption{In contrast to Fig. \ref{fig1}, for the pair of points $a$ and $b$ separated by a distance shorter than $2\tilde z$, the contributions to the differential entropy from the dual situations shown here cancel each other. The left-panel situation gives negative contribution which cancels the positive one from the right-panel one.}\label{fig2}
\end{figure}

     In contrast, for $\ell_{\geq} < 2\tilde z$ we also have two corresponding situations as depicted in Fig. \ref{fig2}. In this case, the situation in the left panel with $a \in I_k - I_{k+1}$ gives the negative contribution to the differential entropy, and cancels the positive contribution from the one in the right panel with $b \in I_{k'} - I_{k'+1}$. This implies that there is no contribution to the differential entropy from the entanglement contour of the pairs separated by a distance larger than $2\tilde z$.

   In summary, the differential entropy constructed by the intervals of length equal to $2\tilde z$ only counts entanglement contour associated with the pairs separated by a distance greater than $2\tilde z$.  The contributing contours measure the long-distance correlation of size beyond $2\tilde z$ (even though not all the long-distance entanglement is captured by the contour) and they can be considered as the manifestation of the cost for performing the task called constrained state merging introduced in \cite{Czech:2014tva}. This task is performed by the operations restricted to within the region of size smaller than $2\tilde z$. 
   
\subsubsection{Integral geometry point of view}

   Now we like to consider how the entanglement contour renormalizes along the RG flow from the integral geometry point of view. We notice that the change of the cutoff line from $z=z_1<\tilde z$ to $z=z_2<\tilde z$ will not affect the measure of the kinematic space (number density of geodesics) at $z=\tilde z$ as long as $z_1, z_2<\tilde z$
\be
\el{reshuflmeasure}
\frac {\p^2 S(u(z_1),v(z_1);z_1)}{\p u \p v} \rmd u(z_1) \wedge \rmd v(z_1) = \frac {\p^2 S(u(z_2),v(z_2);z_2)}{\p u \p v} \rmd u(z_2) \wedge \rmd v(z_2)\,,
\ee
where $u(z),v(z)$ are end-point coordinates defined on different slices. Obviously, the integrated form of \er{reshuflmeasure} also holds. By Crofton's formula, It implies that the differential entropy of some bulk curve (within the region $z > \tilde z$) is invariant as we move the running RG scale from $z=z_1<\tilde z$ to $z=z_2<\tilde z$.

\begin{figure}[tbp]
\centering
\includegraphics[height=55mm]{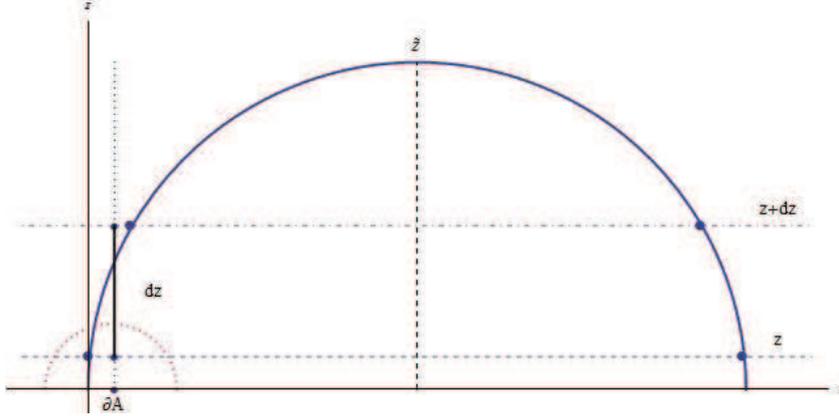}
\caption{The long-distance entanglement is invariant under RG flow but it is reshuffled to shorter scale. Two situations are shown here. The first is the blue geodesic which intersects both surface states at $z$ and $z+dz$. This implies that the entanglement contour for the intersecting pair of points is just reshuffled but not removed. However, the entanglement entropy is reduced (and turned into line element $dz$) as the left end point is dragged across the entangling surface (the intersection points of the vertical bar with the horizontal lines). The second situation is the red-dotted geodesic for which the intersecting points shrink to none as the surface state is pushed up from $z$ to $z+dz$. The first situation is similar to isometry operation in MERA, while the second is to disentangler.}\label{fig3}
\end{figure}

 Moreover, as we can see from Fig.~\ref{fig3} the distance $|u(z)-v(z)|$ between the end points of the same geodesic shrinks along the RG flow. In other words, the same correlation is seen as for shorter scale from the viewpoint of an IR observer.

    To derive some concrete result from the fact of \eq{reshuflmeasure}, we consider renormalization of  the long-distance entanglement contour (of a point) \eq{thetadefinition} (see also \eq{contourpoint}) in homogeneous space, i.e. holographically dual to pure AdS$_3$. For a surface/state $\psi(\tilde z)$, the entanglement entropy between a line element around the point at $(\tilde x, \tilde z)$ and its complement is $\Theta(0, \tilde x; \tilde z)$ as already given by \eq{thetadefinition}.  Due to the assumed homogeneity along x-direction, $\Theta(\ell,\tilde x; z)$ is independent of $\tilde x$  (thus we omit the $\tilde x$ in $\Theta(\ell,\tilde x; z)$ hereafter) so that we can add up the this contour on a horizontal closed line (by imposing periodic boundary condition $\tilde x \sim \tilde x + L_x$) to arrive its length (or the entanglement entropy) $\Theta(0; \tilde z) L_x$.  On the other hand, the length can be seen as the differential entropy for the surface/state  $\psi(z)$, which using \eq{diffIG} and \eq{thetadefinition} is given by $\int \p_{\xD x} S(\xD x) \rmd \tilde x= L_x \Theta(2\sqrt{\tilde z^2-z^2};z)$. The equivalence of the above two different views is guaranteed by the same integral geometric construction and then it yields
\be\el{reshuffle2}
\Theta(2\sqrt{\tilde z^2-z^2};z)=\Theta(0;\tilde z)\;.
\ee    
Eq. \eq{reshuffle2} implies that the long-distance entanglement (more precisely, the part captured by the contour) is reshuffled into shorter scale as one flows the surface state from the UV state $\psi(0)$ to $\psi(z)$.   

    This result of \eq{reshuffle2} can be understood holographically as follows. As shown in Fig. \ref{fig3},  the entanglement contour for the differential entropy associated with the line element around the point $(\tilde x,\tilde z)$ is manifested as the geodesic with $(\tilde x,\tilde z)$ as its highest point.  The locations of the pair of points are then specified by its intersection with the cutoff surface for the UV reference state. As long as $z\le \tilde z$, the entanglement contour remains after the RG step from $\psi(0)$ to $\psi(z)$ but with the separation of the end points change from $2\tilde z$ to $2\sqrt{\tilde z^2-z^2}$.   As the contour of all the other pairs contributed to $\Theta(2\sqrt{\tilde z^2-z^2};z)$ are manifested in a larger geodesic than the above one so that they also remain after the above RG step. Thus the long-distance entanglement contour remains the same after the RG running as long as the running scale $z<\tilde z$.

   In some sense, this reshuffling is similar to the isometry operations of MERA, for which the entanglement of a pair are not removed but rescaled under coarse-graining into that of a new pair. \footnote{On a MERA network, a similar process is realized as transferring of entanglement to the sites on the next layer, which is part of the isometry operation. We note however that the correspondence is not precise because a site in general carries entanglement to both sides and hence the entanglement entropy is not always reduced when the site is dragged across $\p A$.}  This is exactly what we have shown in Fig. \ref{fig3}. On the other hand, there are situations in which the entanglement contour of a pair of points is removed under RG flow, as shown in (the red-dotted geodesic of) Fig.~\ref{fig3}. The entanglement entropy reduced in this case is also turned into (a proportion of) the length of the line element as predicted by the Crofton's formula \eq{Crofton}. This is similar to the disentangler operations in MERA.

In fact, the above entanglement renormalization can be encoded in the following RG equation for the entanglement contour, i.e.,
\be
\el{ERG}
\frac {d }{dz}\xT(2\sqrt{{\tilde z}^2 - z^2};z) = 0\,.
\ee
It is easy to see that $\Theta(\ell_{\geq};z)=2 s(\ell_{\geq};\tilde z)$ given by \eq{contourads3} trivially satisfies \eq{ERG}. Moreover, one can also see that \eq{reshuffle2} is consistent with RG equation \eq{ERG}. \footnote{As the entanglement entropy of finite interval is related to integration of entanglement contour, we think the RG equation of renormalized entanglement entropy in \cite{Jackson:2013eqa} could be related to some integrated form of \eq{ERG}. This issue deserves further study.}

Taking derivative with respect to $z$, we get
\be
\el{ERG2}
  -\p_z \xT(0; z) \rmd z= -2 \p_{\ell_{\geq}} \xT(\ell_{\geq};0)|_{\ell_{\geq}=2z} \rmd z\,. 
\ee
Being positive, the l.h.s. is the change of the total entanglement entropy (of every point) under RG because $\xT(0; z)$ counts the entanglement contribution with any distance from the point $x=\tilde x$. On the other hand, $\xT(\ell_{\geq}; z)$ only counts the ones with distance greater than $\ell_{\geq}$ so that $-\p_{\ell_{\geq}}  \xT (\ell_{\geq};z)$ (also positive) is essentially the entanglement density at a certain length scale $\ell_{\geq}$. This RG equation tells that which portion of the entanglement is removed under each step of the RG flow.

   One can easily generalized the above RG equation for the entanglement contour to more generic background other than the pure AdS$_3$ space. Let $\xD(\tilde z;z)$ denotes the length of the interval spanned by the geodesic and we choose the parameter $\tilde z$ so that $\xD(z;z)=0$, then the RG equation is
\be
\el{ERGG}
\frac {d }{dz}\xT(\xD(\tilde z;z);z) = 0\,,\qquad  \mbox{with} \quad \xD(z;z) = 0\,,
\ee
which implies
\be
\el{ERG2a}
  -\p_z \xT(0; z) \rmd z = -\p_{\ell_{\geq}}  \xT(\ell_{\geq};0) |_{\ell_\geq=\xD(z,0)} \rmd\xD(z;0)\,. 
\ee
Other than a non-trivial Jacobian between $z$ and the length scale on the
boundary, we reach essentially the same conclusion.

Finally we would like to remark on how the RT formula can be realized from the renormalization of entanglement contour.  Both types of LUs (disentangler and isometry) can be visualized as segments of geodesics (red and blue respectively in Fig.~\ref{fig3}) chopped off by the surfaces of RG, which are the analogues of the links (disentangler/isometry) connecting neighboring sites on the MERA network. Taking both types of geodesics (long and short) into account, the Crofton's formula \eq{Crofton} turns into the familiar form in MERA
\be
\label{MERART}
\zt{length of curve } \xg = \# (\zt{LUs cut})\,.
\ee
Moreover, the entanglement entropy of an interval $A$ is bounded by the number
of cuts (LUs that reduce the density matrix $\rho_A$ to the one with no entanglement)
\be
S_A \le \# (\zt{LUs cut}) = \zt{length of curve } \xg \zt{ ending on
}\p A,
\ee
which is precisely the MERA version of RT formula. 

To summarize, the entanglement structure of a quantum state is depicted by the collection
of entanglement contours labeled by pairs of points on the boundary. As RG flow
progresses, the contour of pairs shorter than the cutoff are completely removed and they are no longer relevant for the entanglement of the new state $\psi(z)$ (or its corresponding geometry) because these pairs correspond to the entanglements over scales shorter than the cutoff. All the entanglement contours over longer scales remain intact and are simply relabeled. Both types of operations are local. 

\section{ Generalization to higher dimensions}
\label{sec:IGHigherD}
 We have demonstrated that, using AdS$_3$ as an example, the connection between
entanglement and geometry becomes manifest in the framework of integral geometry. In this section, we would like to generalize this construction to (the $d$-dimensional spatial slice of) $d+1$-dimensional stationary spacetime. 

\subsection{Kinematic space}\el{subsection:inv-K}
The key observation in 3D case we have is that the volume measure of the auxiliary kinematic space of all geodesics is tied to the entanglement at a particular scale. Furthermore, this volume measure can be understood as the entanglement contour measuring the amount of entanglement between two boundary end points of a geodesic. Instead of using geodesics, in higher dimensions one can measure the area (corresponding to holographic entanglement entropy or differential entropy by RT formula) of a co-dimension two surface \footnote{To avoid confusion we always define the co-dimension of an object in the bulk with respect to the $d+1$-dimensional spacetime.} by using the $r$-dimensional ($r\ge 1$) planes via the formula in integral geometry. However, for simplicity and inspired by our construction in 3D case, we will stick to $r=1$ case for which the useful formula of integral geometry is a special case of \eq{IntegralMeasure} (with $r=1$ and $q=d-1$), i.e.,  
\be
\el{GenCrofton}
\int_{M^{d-1} \cap L_1 \ne \emptyset} N (M^{d-1} \cap L_1)\; \xe_\CK =\frac{O_{d}}{O_1} \frac{\xs_{d-1} (M^{d-1})}{4G}\,,
\ee
where $M^{d-1}$ is a $(d-1)$-dimensional target hypersurface of volume $\xs_{d-1} (M^{d-1})$ and $L_1$ are a set of geodesics with the kinematic space measure $\xe_\CK$, each of which intersects $M^{d-1}$ at $N (M^{d-1} \cap L_1 )$ points. As usual we normalize the measure by multiplying $(4G)^{-1}$ so that the integral becomes dimensionless and corresponds to entanglement (or differential) entropy. 

  It is known \cite{santalo,Santalo:1952} that the measure $\xe_\CK$ for geodesics (\ie $r=1$) on a general Riemannian manifold $\CM$ \footnote{The precise criteria the Riemannian manifold has to meet so that it is uniquely determined by the boundary-to-boundary distances remains unknown (see \eg \cite{Porrati:2003na,pestov:2005}). Detailed discussions are beyond the scope of the current paper. We will simply assume $\CM$ has this property of boundary distance rigidity.} can be constructed as follows (see \eg \cite{santalo} for more details). We consider a $d$-dimensional Euclidean Riemannian manifold $\CM$, which is endowed with a metric $g_{\mu\nu}$. The phase space ${\cal X}$ of all geodesics is then a $(2d-1)$-dimensional space with coordinates $(x^\mu, p_\mu)$. The conjugate momentum is related to the velocity $\dot x^\mu$ (dot for derivative with respect to the parameter $\tau$ of each geodesic) by $p_\mu = g_{\mu\nu} \dot x^\nu/|\dot x|$ and is subject to the constraint $p^2 = 1$. The geodesics as integral curves provide a foliation $F_G$ of ${\cal X}$. 
We have the following $(2d-2)$-form
\be
\el{invform}
\xe_{\cal X} = \frac 1 {4G}\lb\sum_{\mu=1}^d \rmd p_\mu \wedge \rmd x^\mu \rb^{d-1}\,,
\ee
whose restriction on ${\cal X}/F_G$ then gives the measure $\xe_\CK$. 

We can choose to parameterize a geodesic using its end points $(\vec x_1, \vec x_2)$ on a slice (with coordinates $x^i\; (i=1,\dots d-1)$), which is a member of the family of convex hypersurfaces foliating $\CM$. The slices also correspond to dual surface states at different RG scales as parametrized by the radial coordinate $x^d$, which can also be treated as the parameter of the geodesic. With the help of action-angle relation of Hamilton-Jacobi argument, i.e., $p_i = \p_i S$, the measure then becomes
\be
\label{generalmeasure}
\xe_\CK = \frac 1 {4G}\det \lsb \frac {\p^2 S(\vec x_1, \vec x_2) }{\p \vec x_1 \p \vec x_2}\rsb \prod_{i=1}^{d-1} \rmd x^i_2 \wedge \rmd x^i_1  \,,
\ee
where we still use $S$ to denote the length of geodesic but generically it no longer has direct connection with entanglement entropy. One can easily see that for $d=2$, the measure \eq{generalmeasure} reduces to \eq{metrickinematic} when the slice is the boundary. Volume integral in the kinematic space is independent of the parameter $\tau$, i.e., 
\ba
\int \p_\tau \xe_\CK & = & \frac {d-1} {4G} \oint \p_\tau \sum_{\mu=1}^d p_\mu \rmd x^\mu \lb \sum_{\mu=1}^d \rmd p_\mu \wedge \rmd x^\mu\rb^{d-2} \nn
& = & \frac {d-1} {4G}\oint d \CL \lb\sum_{\mu=1}^d \rmd p_\mu \wedge \rmd x^\mu\rb^{d-2} = 0\,,
\ea
where $\CL:= \sqrt{g_{\mu\nu} \dot x^\mu \dot x^\nu}$. So we are free to choose any slice, which then justifies the use of $S(u,v;\tilde z)$ in \eq{SeeDiff} and \eq{Bellpair}.

 With the generic measure \eq{invform}, one can show that the Crofton's formula \eq{GenCrofton} remains valid for a general Riemannian manifold (see \eg \cite{santalo}). Here is a sketch of the proof. Let us first consider an infinitesimal area element $\rmd \xs$ on the hypersurface $M^{d-1}$. A local coordinate is chosen so that $M^{d-1}$ is parameterized by $y^d = 0$. We have additional freedom to take $\rmd \xs$ to $y^\mu = 0$ and diagonalize the metric at this point 
\[\rmd s^2 = \sum_{\mu=1}^d g_{\mu\mu} \rmd y^\mu{}^2 \]
We then make use of \eq{generalmeasure} and choose $y^{1, \dots, d-1}$ as the position coordinates for the geodesics. The conjugate momentum can be parameterized using the angle $\xa^\mu$ between the momentum and the vielbein,
\[
 \xa^\mu := e^\mu \cdot p \Rightarrow p_\mu = \sqrt{g_{\mu\mu}} \cos \xa^\mu\,.
\]
The measure \eq{invform} (for geodesics passing through neighborhood of $y^\mu = 0$) then becomes
\be
\xe_\CK = \frac 1 {4G}  \prod_{i = 1}^{d-1} \sqrt{g_{ii}} \sin \xa^i \rmd \xa^i \wedge \rmd x^i\,.
\ee
One can easily pick up the area element $\rmd \xs = \prod_{i = 1}^{d-1} \sqrt{g_{ii}} \rmd x^i$. Notice that the area element for a unit $(d-1)$-sphere is 
\[
\xe_{S^{d-1}} = \frac 1 {\cos \xa^d} \prod_{i = 1}^{d-1} \sin \xa^i \rmd \xa^i\,,
\]
and therefore $\xe_\CK$ can be written as
\be
\xe_\CK = \frac 1 {4G} \cos \xa^d \xe_{S^{d-1}}\wedge\rmd \xs\,.
\ee
To get the area element of $\rmd \xs$, geodesic along every direction needs to be counted (to be precise, we only integrate over half of the solid angle $\cos \xa^d > 0$ to avoid over counting). The angular integration gives a factor of 
\[\int \cos \xa^d \xe_{S^{d-1}} = \frac {O_d}{O_1}\,.\]
As we can see, the Crofton's formula \eq{GenCrofton} is then correctly reproduced. We would like to remind the reader that the results above only applies to the case of $r=1$ and construction for generic $r$ is little known.
 
   As discussed previously for 3D $(d=2)$ case \eq{contourpoint}, the holographic entanglement entropy of an infinitesimal area element at a point $A$ can be expressed in terms of an integral over the geodesics with one of the end points at $A$.  Thus this volume form can be thought as  the entanglement contour specified by the end points of the geodesic.  However, this entanglement contour lacks precise definition in the context of field theory unlike its 3D counterpart.  The length of a geodesic is related to the two-point functions of heavy operators and hence is a measure for the correlation on the field theory side. However, to our best knowledge its precise connection with entanglement entropy remains unclear.  Instead, we can treat it as the holographic definition waiting for the field theory verification.

\subsection{Bulk line elements from intersecting geodesics} 
  
 As in the 3D case, we will now explicitly show how bulk line element can be obtained from the measure \eq{IntegralG} of integral geometry. For simplicity we will only consider the bulk space of homogeneity and isotropy. We would like to emphasize that the Crofton's formula applies to more general spaces but the general prescription to reconstruct the metric from co-dimension two surface is unknown to us. As before let us work in Poincare coordinate, in which the metric takes the form,
\be
\rmd s^2=-\tilde f(z) \rmd t^2+\sum_{i=1}^{d-1}g(z) (\rmd x^i)^2+ g(z)f(z)\rmd z^2\,.
\label{perturbgeneral}
\ee
As in a $2$-body system we can go from the end point coordinates $(\vec x_1, \vec x_2)$ to the mid-point
$\vec{x}:=\frac 1 2 (\vec x_1+\vec x_2)$ and their separation $\xD \vec x:= \vec x_1 - \vec x_2$, the latter of which can in turn be expressed in terms of the distance $2 r := |\vec x_1 - \vec x_2|$ and the orientation $\Omega$. The measure in the kinematic space takes the following form
\be
\el{IntegralG}
\xe_\CK = \frac 1 {4G}\xo(\vec x,\xa,\xO)\, \xe_{B_{d}} \wedge dr \wedge d\xO\,,
\ee
where $\xe_{B_{d}}$ is the volume measure of the boundary, i.e., for $\vec x$. 

  Due to the homogeneity and isotropy, the measure of the kinematic space can be characterized only by the separation $2 r$ of the end points of the geodesic, or equivalently and more conveniently by its height at $z=z_*$ (along the radial direction), i.e., 
\be
\label{measure2body}
\xe_\CK = \frac 1 {4G}\frac {\omega_h (z_*)}{O_{d-2}}\; \xe_{B_{d}} \wedge \rmd z_* \wedge d\xO\,.
\ee
Note that, the relation between $r$ and $z_*$ can be shown to be
\be
\el{heightgeo}
r =  \sqrt{g(z_*)}\int_{\tilde z}^{z_*}\rmd z\frac{\sqrt{f(z)}}{\sqrt{g(z)-g(z_*)}}\;.
\ee


It is not difficult to see that due to the symmetry the matrix in \eq{generalmeasure} satisfies
\be \bay{cc} \frac {\p^2 S(r) }{\p x_1^1 \p x_2^2} & \dots \nn \dots & \dots \eay \bay{ccc} \frac {\p x_2^1} {\p r}& \frac {\p x_2^1} {\p \xt} & \dots \nn \frac {\p x_2^2} {\p r}& \frac {\p x_2^2} {\p \xt} & \dots \nn \dots & \dots & \dots \eay = \bay{ccc} -\frac {\p^2 S(r)}{\p r^2} \frac {\p x_2^1} {\p r}& \frac 1 r \frac {\p S(r)}{\p r} \frac {\p x_2^1} {\p \xt} & \dots \nn -\frac {\p^2 S(r)}{\p r^2} \frac {\p x_2^2} {\p r}& \frac 1 r  \frac {\p S(r)}{\p r} \frac {\p x_2^2} {\p \xt} & \dots \nn \dots & \dots & \dots \eay\,, \ee
where $r,\xt,\dots$ are the polar coordinates for $\xD \vec x$ and hence
\be
\det \lsb\frac {\p^2 S(r) }{\p \vec x_1 \p \vec x_2}\rsb = -\frac {\p^2 S(r)}{\p r^2} \lb \frac 1 r \frac {\p S(r)}{\p r} \rb^{d-2}\,.
\ee
Using the facts that $\p_r S(r) =2\sqrt {g(z_*)}$ and $O_d/O_1 = O_{d-2}/(d-1)$, we can rewrite \eq{generalmeasure} into the form of \eq{measure2body} to extract $\omega_h(z)$ as given by
\be
\el{vol_kin2}
\omega_h(z) =-\frac {O_d}{O_1} \frac {d-1} 4 [g(z)]^{\frac {d-3} 2} \frac {d g(z)}{d z}\,.
\ee
Notice that a factor of $1/2$ is put in so that we can sum over all the orientation $\Omega$ in applying \eq{GenCrofton}.

\begin{figure}[tbp]
\begin{minipage}[t]{0.5\textwidth}
\includegraphics[height=45mm]{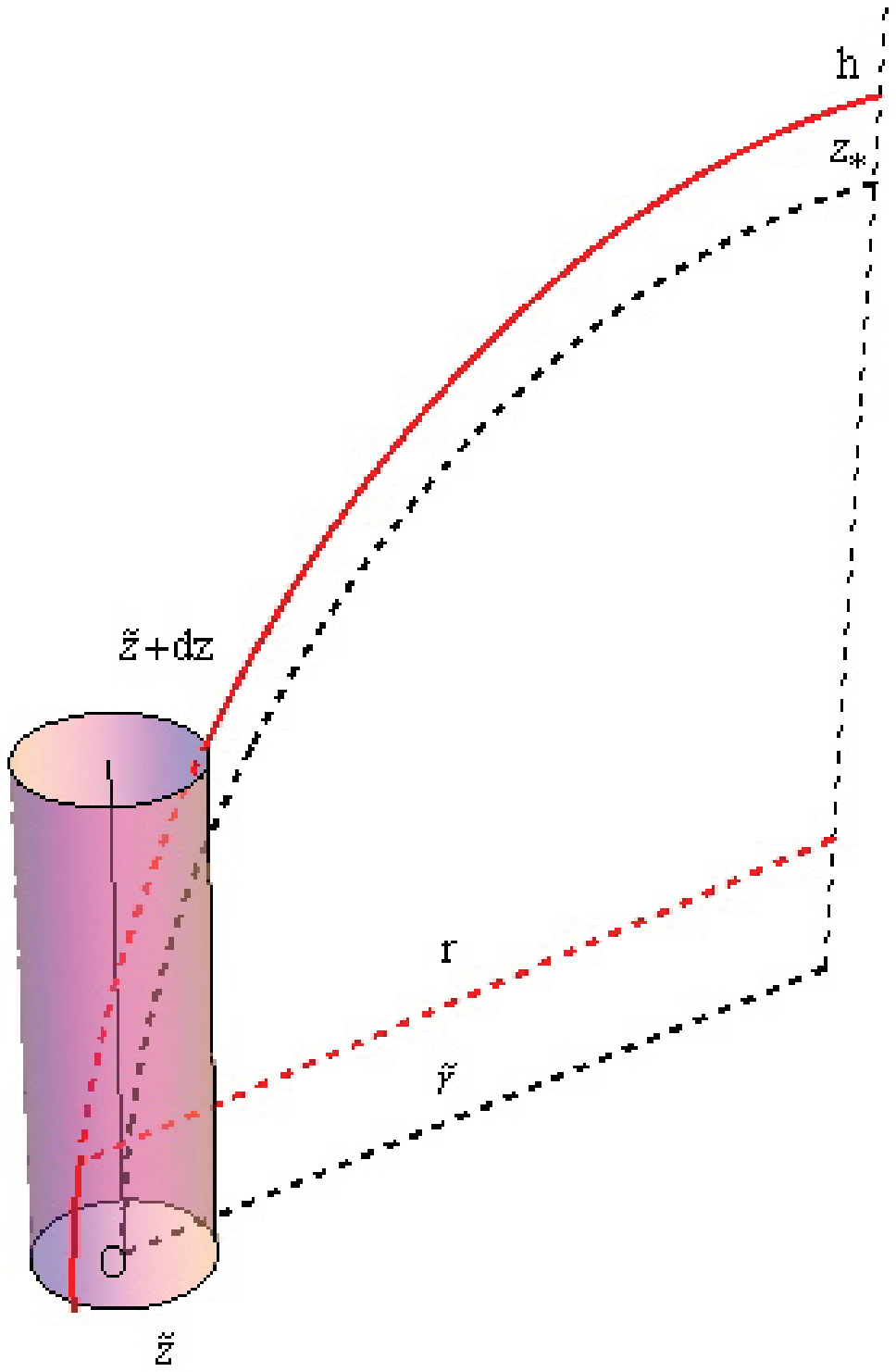}
\end{minipage}
\begin{minipage}[t]{0.5\textwidth}
\includegraphics[height=45mm]{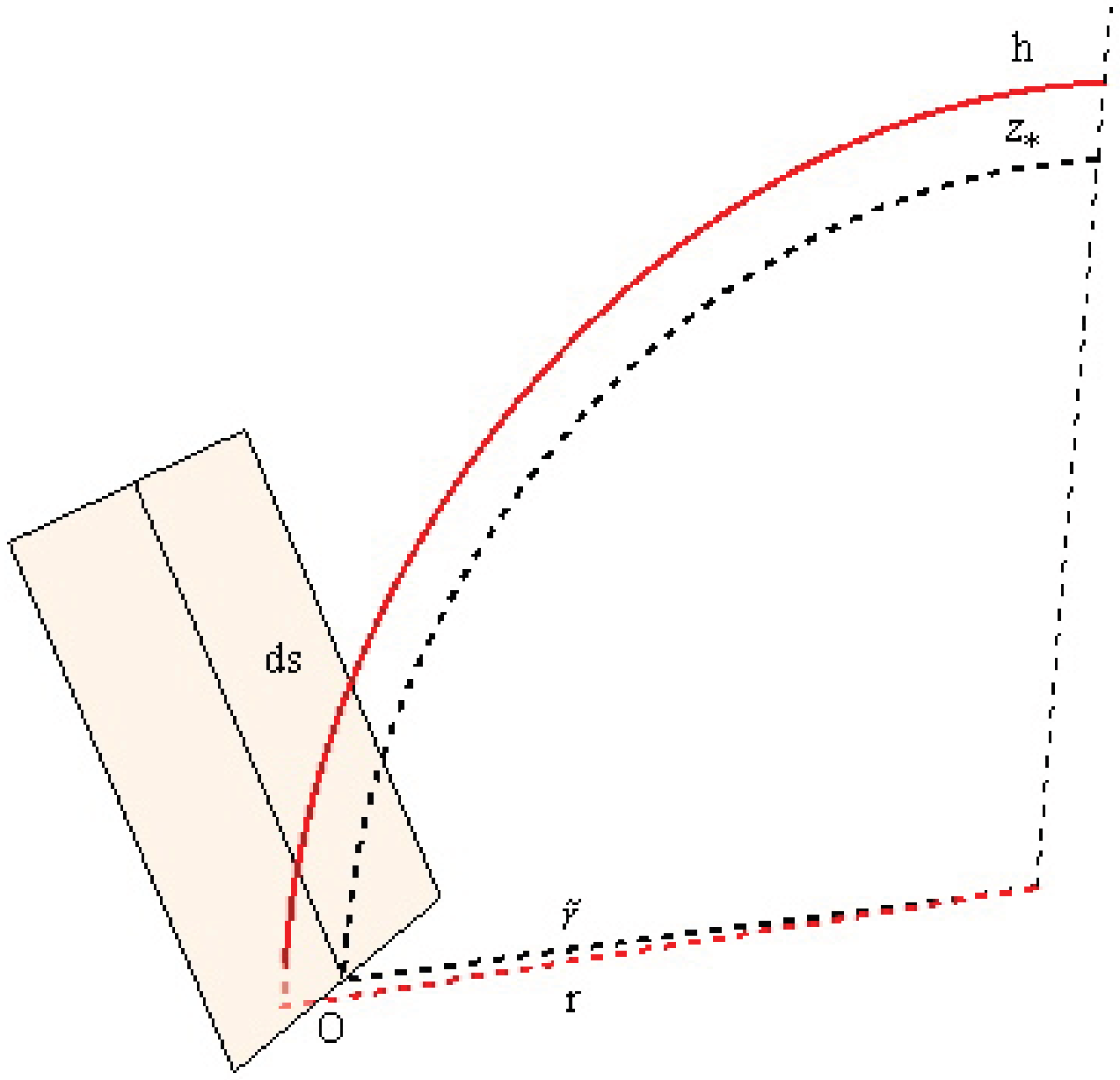}
\end{minipage}
\caption{Various coordinates used in \eq{line-element-hd} (left) and \eq{generallineE} (right). In the former case, we consider a cylinder surrounding the line element from $(0,\tilde z)$ to $(0,\tilde z+dz)$. The size of the cylinder has been exaggerated for artistic reason. The origin $O$ of the polar coordinate lies at $(0, \tilde z)$. The horizontal distance between $O$ and the mid-point is denoted by $\tilde r$, which can be mapped to a new variable $z_*$ using the black geodesic. Geodesics with the same $\tilde r$ yet different heights hit the cylinder at various points. For example, the red geodesic hits the cylinder at $z = z',\, z' \in (\tilde z, \tilde z+ dz)$ and we can define the horizontal distance $r$ from the intersection point to the mid-point (red dashed line). In the latter case, the only difference is that now $r$ is the distance between two points on the slice $z = \tilde z$.}\label{figH}
\end{figure}

   We are ready to calculate the line element $\rmd s$ along the radial direction by applying \eq{GenCrofton} similar to AdS$_3$ case except that $M^{d-1}$ is not a line but should be approximated by an infinitesimal cylinder of length $ds$ and spherical cross-section of radius $\xd$.  We note that for a cylinder of infinitesimal size, no geodesic can intersect it more than twice. For convenience, let us introduce a polar coordinate $(\tilde r, \tilde \Omega)$ on the horizontal surface $\CH$ at $z=\tilde z$ with the starting point of the line element as its origin (\ie the line element is stretching from $(0, \tilde z)$ to $(0, \tilde z +d z)$). We thus label by $\tilde r$ the distance between the origin of the polar coordinate and the projection of the mid-point of the geodesic to $\CH$. On the other hand, the horizontal distance between the lower intersection point of the geodesic with the cylinder and the mid-point is just $r$, which is related to the height of the mid-point $z_*$ by \eq{heightgeo}. A geodesic has to hit the cylinder somewhere ($z'$) between $\tilde z$ and $\tilde z + d z$ to contribute and therefore its height has to lie in a certain range (determined by \eq{heightgeo}). To avoid confusion, we denote the height of a geodesic as $h$ (instead of $z_*$), which is related to $r, z'$ by \eq{heightgeo}. See the left panel of Fig.~\ref{figH} for the graphic specifications. 
With all these specifications of coordinate variables, the detailed steps to compute the area of the cylindrical line element are shown as follows:

\ba
& & 4G \int_{(S_{d-2} \times \rmd s) \cap L_1 \ne \emptyset} 2 \xe_\CK  = 2\int_0^\infty \tilde r^{d-2} \rmd \tilde r\int \rmd \tilde \xO\,\int_{h(r,\tilde z)}^{h(r,\tilde z+\rmd z)}\int\omega_h(h)
\rmd h  \frac{\rmd \xO}{O_{d-2}}\nn
& =& -\frac {(d-1)O_{d-3}O_d} {(d-2)O_{d-2}O_1} \xs_{d-2}(\xd)  \rmd z \int_{g(\tilde z)}^0 \frac{\p z_*}{\p \tilde z}\at{r} \frac {\p r}{\p z_*}\at{\tilde z} g(z_*)^{\frac {d-3} 2} \rmd g(z_*) \nn
& = & \frac {(d-1)O_{d-3}O_d} {(d-2)O_{d-2}O_1}\xs_{d-2}(\xd)\rmd z  \int_0^{g(\tilde z)}\frac{\sqrt{f(\tilde z)}}{\sqrt{g(\tilde z)-y}} y^{\frac {d-2} 2} \rmd y \nn
& = & \frac {(d-1)O_{d-3}O_d} {(d-2)O_{d-2}O_{d-1}}\frac {O_d}{O_1}\sqrt{f(\tilde z)}g(\tilde z)^{\frac {d-1} 2}\xs_{d-2}(\xd)\rmd z = \frac {O_d}{O_1}\sqrt{f(\tilde z)}g(\tilde z)^{\frac {d-1} 2}\xs_{d-2}(\xd)\rmd z\,. \label{line-element-hd}
\ea
In the above, the orientation of the geodesic is also limited to a small solid angle of the size so that
\be
\el{solidangle}
\int \rmd \xO = \frac 2 {d-2} \lb\frac \xd {\tilde r} \rb^{d-2} O_{d-3} =  \frac {2\xs_{d-2}(\xd)} {\tilde r^{d-2}} \frac {O_{d-3}} {(d-2)O_{d-2}}\,,
\ee
where $\xs_{d-2}(\xd)$ is the area of the spherical cap.  Notice that by definition the geodesics are oriented so that integration of solid angle $\xO$ is over the whole range, which leads to the factor of $2$ in \eq{solidangle} as both orientations contribute.  These facts are used to arrive the 3rd line of \eq{line-element-hd}. Besides this, we also need to perform the integration of $h$ (to get $\p_{\tilde z} z_*|_r \rmd z$) and then change the integration variable $\tilde r$ to a new variable $z_*$ using \eq{heightgeo}, i.e., we use the chain rule:
\be
\el{chainrule}
\frac{\p z_*}{\p \tilde z}\at{r} \frac {\p r}{\p z_*}\at{\tilde z}=-\frac {\p r}{\p \tilde z}\at{z_*}  =\sqrt{g(z_*)}\frac{\sqrt{f(\tilde z)}}{\sqrt{g(\tilde z)-g(z_*)}}\,.
\ee
Note that the difference between $\tilde r$ and $r$ is $\CO(\xd)$ so that it will contribute to the $\CO(\xd^2)$ order and can be neglected for the area. Moreover, we should also express $\xo_h \rmd z_*$ in terms of $\rmd g(z_*)$ using \eq{vol_kin2}. We can then read off the line element $\rmd s$ from the area,
\be
\zt{vol}_{S_{d-2} \times ds} = \sqrt{f(\tilde z)g(\tilde z)}\rmd z \times g(\tilde z)^{\frac {d-2} 2} \xd_{d-2}
(\xd) \Rightarrow \rmd s = \sqrt{f(\tilde z)g(\tilde z)}\rmd z\,,
\ee
which is exactly what follows from the metric.

We notice that the derivation above is completely general and is independent of the precise forms of the metric and geodesic (we do assume homogeneity and isotropy). After the cancellation of $r^{d-2}$ from \eq{solidangle}, the specific form of $r(h,\tilde z)$ is no longer needed.

Generic line element can be computed in a similar way. As in the 3D case, we can parameterize the line element using $z$ coordinate. Without lost of generality, we can place the line element along $x^1$-direction \ie $\rmd x^1 = k \rmd z$. In this case, it is more convenient to consider an infinitesimal {\it strip} aligned along the line element with the widths of $\xd$ in the transverse dimensions ($x^2,\dots x^{d-1}$), i.e.,  $(x^1,x^2,\cdots, x^{d-1},z)=(k dz, \xd,\dots, \xd, dz)$, and hence the cross section (surface orthogonal to the line element) of this strip is $\xs_{d-2}(\xd)$. The specifications of the coordinate variables are given in the right panel of Fig.~\ref{figH}, following that the area of the strip line element can be again calculated by using \eq{GenCrofton},
\ba
4G\int_{B_{d-1} \cap L_1 \ne \emptyset} \; \xe_\CK & =&   
\int_0^\infty \tilde r^{d-2} \rmd \tilde r \int \rmd \tilde \xO\,\int\int\omega_h \rmd h  \frac{\rmd \xO}{O_{d-2}}\nn 
& =& \int_0^\infty \tilde r^{d-2} \rmd \tilde r \int \rmd \tilde \xO\,\int \int\omega_h \frac{\p z_*}{\p r}\at{\tilde z}\rmd r\frac{\rmd \xO}{O_{d-2}} \nn
& = & \frac {(d-1)O_d} {2O_{d-2}O_1}\xs_{d-2}(\xd)\rmd
z\int_0^{g(\tilde z)} \int\left|k y^{\frac {d-3} 2}-\cos\xt\frac{\sqrt{f(\tilde z)}}{\sqrt{g(\tilde z)-y}} y^{\frac {d-2} 2} \right| \rmd y \rmd\tilde \xO\nn
& = &  \frac {O_d}{O_1}g(\tilde z)^{\frac {d-1} 2}\xs_{d-2}(\xd) \sqrt{k^2+ f(\tilde z)}\rmd
z \, \label{generallineE}.
\ea
One can then easily pick up the line element $\rmd s = \sqrt{k^2+ f(\tilde z)}\rmd z$ from the area. In the third line of \eq{generallineE}, we rewrite the angular volume form in terms of Cartesian coordinates

\be
\el{volumetrans}
r^{d-2} \rmd r \rmd \xO = 2\rmd x^1 \dots \rmd x^{d-1}- \frac {2x^1}
r \frac{\p r}{\p \tilde z}\at{z_*} \rmd z  \rmd x^2 \dots \rmd x^{d-1}\,,
\ee
where $r$ is now the horizontal distance from the midpoint of the geodesic intersecting the strip to its end point on $\CH$, and $x_1=r \cos \xt$. At the lowest order (in $dz$ and $\xd$) we have $r = \tilde r$. The $z$-component in the line element provides the second term in \eq{volumetrans}. The factor of $2$ is due to orientation as explained above. We note that the integrand flips sign where the contributing region in the kinematic space pinches off. This can be remedied by manually putting in the absolute sign. To evaluate the integral in the third line, we can perform the following
change of variable, $y = {g(\tilde z)}/{(1 + w^2)}$ and the integral in \eq{generallineE} then becomes
\ba
& & \int_0^{g(\tilde z)}\int\left|k y^{\frac {d-3} 2}-\cos\xt\frac{\sqrt{f(\tilde z)}}{\sqrt{g(\tilde z)-y}} y^{\frac {d-2} 2} \right| \rmd y \rmd \tilde \xO \nn
&= &2g(\tilde z)^{\frac{d-3}{2}}\int_0^\infty \int\left| \frac {k w} {\left(w^2+1\right)^{\frac{d+1}{2}}}-\frac {\cos \xt \sqrt {f(\tilde z)}} {\left(w^2+1\right)^{\frac{d+1}{2}}}\right| \rmd w \rmd\tilde \xO\nn
& = & 2g(\tilde z)^{\frac{d-3}{2}}O_{d-3}\lsb\int_0^\infty \int_{\frac \pi 2}^{\pi}\lb \frac {k w} {\left(w^2+1\right)^{\frac{d+1}{2}}}-\frac {\cos \xt \sqrt {f(\tilde z)}} {\left(w^2+1\right)^{\frac{d+1}{2}}}\rb \sin^{d-3}\xt\rmd w \rmd \xt \rc\nn
& & \lc +\lb \int^\infty_{\frac {\cos \xt \sqrt {f(\tilde z)}} k}-\int_0^{\frac {\cos \xt \sqrt {f(\tilde z)}} k}\rb \int_0^{\frac \pi 2}\lb \frac {k w} {\left(w^2+1\right)^{\frac{d+1}{2}}}-\frac {\cos \xt \sqrt {f(\tilde z)}} {\left(w^2+1\right)^{\frac{d+1}{2}}}\rb \sin^{d-3}\xt\rmd w \rmd \xt\rsb\nn
& = & g(\tilde z)^{\frac{d-3}{2}}O_{d-3}\lsb \frac{\sqrt{\pi }\,  \Gamma \left(\frac{d-2}{2}\right) \left(2 \sqrt{f(\tilde z)+k^2}-\sqrt {f(\tilde z)}-k\right)}{2 \Gamma \left(\frac{d+1}{2}\right)}+\frac{\sqrt{\pi }\, \Gamma \left(\frac{d-2}{2}\right) (\sqrt {f(\tilde z)}+k)}{2 \Gamma \left(\frac{d+1}{2}\right)}\rsb\nn
& = & \frac {O_{d}} \pi g(\tilde z)^{\frac{d-3}{2}}\sqrt{f(\tilde z)+k^2}\,.
\ea

 We have shown that the the metric \eq{perturbgeneral} other than pure AdS space can be reconstructed using the generalized Crofton's formula \eq{GenCrofton}. In arriving the line elements we have used the explicit form \eq{heightgeo} of the geodesic to determine the point curves, i.e., one should know the bulk metric in advance to determine the explicit form of geodesic. Thus, our results can only been seen as a consistency check of the integral method to generic space, but not a priori derivation of the line element. However, as shown in \cite{Czech:2015qta},  the point curves can be constructed iteratively purely in kinematic space without knowing bulk metric and the explicit from of geodesic. For completeness, the procedure of the construction is briefly outlined below.
 
  We assume the point curves for all the points with radial coordinate up to $z = \tilde z$ are known. Each point $A$ corresponds to a $d-1$ dimensional subspace $p_A$ in the kinematic space, which we can parameterize as $v_A{}^i(u)$ using the end points $u^i,v_A{}^i$ ($i=1, \dots d-1$). As we have shown, all the line elements and therefore the metric in this region ($z \le \tilde z$) can be reconstructed. The point is then to compute the quantity $\cd_{x_A} v_A{}^i$ which are the shift functions that determine the new point curve. We use $x_A$ to denote the coordinates of the bulk point $A$ and $v_A{}^i$ for the point curve. For every component $i$, we essentially need to determine the 1-forms $\cd_{x_A} v_A{}^i$. Some of its components $t_j \cdot \cd v_A{}^i$, where $t_i$ are the tangent vectors along the surface $z=\tilde z$, are already known. What remains to be found is the component $n \cdot \cd v_A{}^i$ along the orthogonal direction. Moreover, the component $k \cdot \cd v_A{}^i$ along the unit tangent vector $k^i$ of the geodesic $v_A{}^i(u)$ always vanishes as the same geodesic (starting from end point $u$) passes through every point on itself. Angles between $k$ and $t_i, n$ can be computed from $t_i \cdot \cd S$ ($S$ being the length of geodesic which follows from \eq{GenCrofton}). As a result $n \cdot \cd v_A{}^i$ can be obtained by solving $k \cdot \cd v_A{}^i = 0$. Let us take the space \eq{perturbgeneral} as a specific example. As we can see from \eq{chainrule}, the vector $\p_{\tilde z} z_*|_{r}$ we are after can be obtained from the known point curve $\p_{z_*} r|_{\tilde z}$ and the conjugate momentum $\p_{\tilde z} r|_{z_*}$ (\ie $\cd l$, the gradient of the length of geodesic with respect to the end point).

\subsection{Holographic entanglement entropy and its renormalization from integral geometry}

The formula \eq{GenCrofton} can also be applied to a minimal surface whose
area is the holographic dual of the entanglement entropy according to Ryu-Takayanagi formula. Here we will only consider the entanglement entropy in vacuum across a spherical surface. This sphere of radius $R$ is the boundary of a disk $D$ and the surface in bulk (pure AdS) is denoted by $\xg_D$. In this case, only geodesics with either one of the end points inside the disk
$D$ contribute to the integral. For convenience here we switch to the coordinate of end points in which the measure is given by the distance between the two points
\be
\xe_\CK = \frac {L^{d-1}}{4G}\frac {\omega'(|\vec {\tilde x}-\vec x|)}{O_{d-2}}\;  \rmd^{d-1} \vec {\tilde x} \wedge \rmd^{d-1} \vec x ,\quad \xo' = 2^{d-1} (d-1) \frac {O_d}{O_1} |\vec {\tilde x} - \vec x|^{2-2d}\,.
\ee
The area can be expressed in the following form
\ba
\frac {\xs(\xg_D)} {4G} & = & \frac {O_1}{O_d} \int_{S_{d-1} \cap L_r \ne \emptyset}\xe_\CK = \frac {L^{d-1}}{4G}\int_0^R\tilde r^{d-2} \rmd \tilde r\int \rmd \tilde \xO\,\int_R^\infty\int\xo'\, r^{d-2}\rmd r \frac{\rmd \xO}{O_{d-2}}\nn
& =& \frac {L^{d-1}}{4G}\int_0^\infty \tilde r^{d-2}\rmd \tilde r\,\int_R^\infty \int2^{d-1} (d-1) (\tilde r^2-2 \tilde r r \cos (\theta )+r^2)^{1- d} r^{d-2}\rmd r \rmd \xO\nn
& =& \frac {L^{d-1}}{4G}\int_0^R \tilde  r^{d-2}\rmd \tilde r\,\int_R^\infty r^{d-2}\rmd r \frac{4^{d-1} (d-1) \pi ^{\frac d 2} \csc \frac{d \pi }{2} \left(\tilde r^2+r^2\right)}{\Gamma \left(1-\frac{d}{2}\right) \Gamma (d-1) \left(r^2 - \tilde r^2\right)^{d}}\,.
\ea
As expected, the area is divergent (as the limits of the integral approach
$R$) and we need to impose some kind of cutoff $\xd$
\ba
\el{RTIG}
\frac {\xs(\xg_D)} {4G} & =& \frac {L^{d-1}}{4G}\int_0^{R- R \xd} \tilde r^{d-2}\rmd \tilde
r\,\int_{R+ R \xd}^\infty r^{d-2}\rmd r \frac{4^{d-1} (d-1) \pi ^{\frac d 2} \csc \frac{d \pi }{2} \left(\tilde r^2+r^2\right) }{\Gamma \left(1-\frac{d}{2}\right) \Gamma (d-1)\left(r^2-\tilde r^2\right)^{d}} \nn
& = & \frac {L^{d-1}}{4G}\int_{\frac{4 \xd }{(\xd +1)^2}}^1 \frac{2^{2 d-3} \pi^{\frac
d 2}   \csc \frac{d \pi }{2}}{\Gamma \left(1-\frac{d}{2}\right) \Gamma (d-1)}\frac {(1-t)^{\frac{d-3}{2}}}{t^{d-1}} \rmd t
\ea
We note that the final result is regularization dependent and the cutoff we use is by no means the same as that used in computing the area of $\xg_D$ (cutoff in radial coordinate). As a result \eq{RTIG} is different from the familiar form \cite{Ryu:2006ef},
\be
\el{RTsph}
\frac {\xs(\xg_D)} {4G} = \frac {O_{d-2} L^{d-1}}{4G} \int_{\xd}^1\frac{\left(1-y^2\right)^{\frac{d-3}{2}}}{y^{d-1}} \rmd
y\,.
\ee

However, it can be shown that \eq{RTIG} has the right divergent behavior and gives the same universal term as \eq{RTsph} in arbitrary dimensions 
\be
\frac {\xs(\xg_D)} {4G} = \frac {a_{d-2}} {\xd^{d-2}} \dots + \frac {a_{d-2-2k}} {\xd^{d-2-2k}} \dots + \left\{ \begin{array}{ll} a_0 \log \xd & \quad, \zt{d even}\\ a_0 & \quad, \zt{d odd} \end{array}\rc\,,\quad k \in \zt{integers}\,,
\ee
where
\be
a_0 =\frac {L^{d-1}}{4G} \times \left\{ \begin{array}{ll} \frac{(-1)^{\frac d 2} O_{d-2} (d-3)\text{!!} }{(d-2)\text{!!}} & \quad, \zt{d even} \\ \pi ^{\frac{d}{2}-1} \Gamma \left(1-\frac{d}{2}\right) & \quad, \zt{d odd} \end{array}\rc\,.
\ee
For example, when $d=4$ \eq{RTIG} gives

\be
\frac {\xs(\xg_{D_3})} {4G} =\frac {L^{d-1}}{4G}\lb \frac{\pi }{2 \xd ^2}+2 \pi  \log \xd  + \dots \rb\,,
\ee
whose universal term agrees with the one from \eq{RTsph}.

  In a higher dimensional homogeneous space, we can add up the pairs separated by a scale greater than $\ell_\ge$ and define the long-distance entanglement contour $\xT(\ell_\ge,\vec x; \tilde z)$ as in \eq{thetadefinition}
\be
\el{thetahigherD}
\xT(\ell_\ge,\vec x; \tilde z)\rmd^{d-1} \vec x:=\int_{\ell_\ge}^\infty \int\bar{s}(r,\vec x;\tilde z) r^{d-2} \rmd r \rmd \xO\,.
\ee
As before we can consider the case when $M^{d-1}$ is a horizontal surface $\CH$ at constant $z = \tilde z$. Only geodesics of the height $h$ greater than $\tilde z$ can intersect the surface and the area $\xs(\CH)$ is given by the volume of the region $h > \tilde z$  in the kinematic space and naturally the volume form $\xo(x,h,\xO)$ can be understood as contribution from entanglement contour above the particular scale $h$ ($h >\tilde z$).
It is easy to see that regardless of dimensions, the long-distance entanglement contour is invariant under the RG flow. Moreover, evolution of the entanglement contour under the RG flow is the same as described previously in \S~\ref{subsec:econtour} since it follows from how the horizontal slices cut the geodesics. Each pair associated with some entanglement contour contracts along the flow until the contour is eventually removed from the spectrum. In other words, we have the same RG equations \eq{ERGG} \eq{ERG2a} as before and the picture of LUs acting as disentangler and isometry on the entanglement contours also remains valid. The generalized Crofton's formula then implies that the area of a (co-dimension two) surface is given by the number of LUs it cuts, which is the higher dimensional generalization of \eq{MERART}.

\section{Conclusion}
\label{sec:conclusion}
In this work we propose a new physical interpretation for the measure in the kinematic space of integral geometry as the entanglement (contour) associated with the two end points of geodesics on the boundary. We then show that along the RG flow the entanglement contour evolves as if under the influence of local unitary operations in MERA (or rather cMERA). Moreover, a RG equation for the long-distance entanglement contour is derived, which could be used as a guide for deriving the bulk dynamics. Our result is the refinement of the AdS/MERA duality by looking into the renormalization of the entanglement contour for a pair of points, which is a new type of measure of spatial entanglement. 

   The central theme of this study is to understand bulk geometry, which is encoded in the kinematic space of the integral geometry, from the entanglement of the boundary field theory, or vice versa.  In AdS$_3$ case the measure of the kinematic space is tied to the entanglement entropy so that our construction of bulk geometry can have a thorough field theory realization.  This is however, not the case in higher dimensional cases as it is not known whether there is entropic interpretation of the kinematic space.  The difficulty also lies in the fact that we are short of field-theoretical tool of calculating the entanglement contour despite of the success of evaluating the entanglement entropy in the past few years.  Therefore, our result though based on holography can be thought as an urge for field theorists to take more seriously this microscopic understanding of entanglement renormalization. 
 
   In our approach the entanglement contour is the physical interpretation of $\xe_\CK$,  which may shed some light on how in general dimensions it can be computed from the quantum entanglement of a field theory state in higher dimensions. Unfortunately, at this point its connection with entanglement entropy remains unclear. We also notice that in general if multi-partite entanglement is involved, the mutual information is monogamous (see \cite{Hayden:2011ag} for relevant discussions)
\[
I(B,A_1)+I(B,A_2) \le I(B,A_1 \cup A_2)\;,
\]
and hence the mutual information is not additive for a general many-body quantum state in contrast to the entanglement contour.   
Thus, this will make the task of finding an additive entanglement density such as the entanglement contour in field theory side more difficult.  

Instead of geodesics, a different option for the building block of bulk geometry is the spherical minimal surfaces. The construction of point curves and distance in pure AdS is straightforward. However, in this case it seems more difficult to find an entropic interpretation for the generalized Crofton's formula because the intersection is a $(d-2)$-dimensional object not a point. 

  One ultimate goal of AdS/MERA duality is to derive the bulk dynamics such as Einstein equation purely from the structure of entanglement renormalization of MERA without any a priori input from the bulk side such as the bulk metric. We hope that our result can shed some light on this issue in a more concrete way in the near future.

\acknowledgments
XH is supported by MoST grant: 103-2811-M-003-024 and FLL is supported by MoST grant:103-2112-M-003 -001 -MY3.

\appendix
\section{Differential entropy in general dimensions}
\label{sec:dereivew}
For reader's convenience, here we collect some background materials on differential entropy in general dimensions (mostly from \cite{Headrick:2014eia, Czech:2014ppa}). The idea of hole-ography is that the area $A$ of a (co-dimension two) bulk surface $\xg_b$ can be constructed from differential entropy $E[I(\xl)]$ via Bekenstein-Hawking entropy formula
\be
\el{BHdiff}
E[I(\xl)] = \frac {A(\xg_b)} {4 G}\,,
\ee
$G$ is the Newton constant in the bulk. Differential entropy is defined using the entropy of a collection of co-dimension one strips $I(\xl)$ (or intervals for 2d field theories) that cover the boundary. All these strips are determined by the minimal surfaces
tangent to the bulk surface. More precisely, we can foliate the bulk surface
$\xg_b$ using co-dimension three slices $s_b(\xl)$ parameterized by $\xl$ and at each slice there is a unique minimal surface $\xG(\xl)$ tangent to
$\xg_b$. The surface $\xG(\xl)$ hits the boundary at two disconnected surfaces $\xl_L(\xl)$ and $\xl_R(\xl)$, which are the end points (boundary) of a strip. Differential entropy is then defined for the strips each of which is specified by the two end points $(\xg_L(\xl), \xg_R(\xl))$
\begin{equation}
\label{GdiffE}
E=\oint \rmd\lambda\,\left.\frac{\p S(\gamma_L(\lambda),\gamma_R(\lambda'))}{\p
\lambda'}\right|_{\lambda'=\lambda}\,.
\end{equation}
where $S$ holographically is computed by the area of minimal surface, \ie saddle point of the action with end points $(\gamma_L,\gamma_R)$,
\be
S = \int \CL(\xg,\dot \xg) \rmd \xs \rmd s =\frac 1 {4G} \int \sqrt{\det (\p_a \xg^\mu \p_b
\xg^\nu  g_{\mu\nu})} \rmd \xs \rmd s\,.
\ee
Here we parameterize the surface using slices labeled by $s$ and $\xs^i$
($i = 1 \dots d-2$) are the coordinates on each slice. The area is expressed
in terms of the determinant of the induced metric where the indices $a,b$
run over $s$ as well as all the $\xs$'s and $\mu,\nu$ run over all $d$
spatial coordinates. Let us give a sketch of proof how the integral \eq{GdiffE} can reproduce the length of the bulk curve. We note that it can be rewritten in the form of action angle variable
\begin{equation}
E=\oint \rmd\lambda\, \xg_R'\frac{\p S(\gamma_L,\gamma_R)}{\p
\xg_R} = \oint \rmd\lambda\, \xG_\mu'p^\mu\at{s_R}\,,\label{EAction}
\end{equation}
the latter of which because of the periodicity is a constant of motion (evolution
with respect to $s$)
\be
\el{Diffbound}
\oint \rmd\xl \frac {d S} {d \xl} =  \oint \rmd \xl \lb \xG_\mu'p^\mu\at{s_2} - \xG_\mu'p^\mu\at{s_1}\rb = 0\,.
\ee
Therefore we can move the point along each geodesic to the contact point
$\xg_b(\xg) = \xG(\xg, s_b)$ and the differential entropy becomes the length of the curve,
\be
E = \oint \rmd\lambda\, \xG_\mu'p^\mu\at{s_b} = \oint \rmd\lambda\, \CL(\xg_b(\xl),\xg_b'(\xl))\,.
\ee

The integration needs to be performed in closed contour so that there is no contribution from boundary term \eq{Diffbound}. However, this constraint can be relaxed. For example, in three dimensions, one can define the integral 
\be f(\xt, \tilde \xt) = \frac 1 2 \int_{\xt}^{\tilde\xt} \frac {d S}{d \xa} \rmd \xt\,, \ee
where the angular coordinate $\xt$ serves the role of $\xl$. The function $\xa(\xt)$ is the half length of the interval centered at $\xt$ and subtended by the geodesic passing through $\tilde \xt$.\footnote{The
radial coordinate $R$ of the point is related to $\xa(\xt), \xt, \tilde \xt$ via eq.\eq{alpha}.} This is essentially the differential entropy for a point, which we will discuss in more details in Appendix~\ref{sec:higherD}. In pure AdS$_3$, it takes the following form 
\be
\label{finitediffEP}
f(\xt, \tilde\theta) = \frac{L}{8G} \log\frac{\sin \big[ \alpha(\xt) + \tilde\theta - \theta\big]}{\sin \big[\alpha(\xt) - \tilde\theta + \theta\big]}\,,
\ee
where $\tilde \xt$ is the angular coordinate of the point. As explained in \cite{Czech:2014ppa},
this quantity is tied to the boundary contribution of differential entropy,
namely that
\be
\el{diffendpoint}
\frac {A} {4 G} = \frac 1 2\int_{\xt(\tilde \xt_i)}^{\xt(\tilde \xt_f)} \frac {d S}{d \xa} \rmd \xt + f(\xt(\tilde \xt_f), \tilde\theta_f) - f(\xt(\tilde \xt_i), \tilde \xt_i)\,.
\ee
In this equation, we relax the constraint in defining differential entropy
and now its range of angular coordinate has end points whose angular coordinates are $\tilde \xt_i$ and $\tilde \xt_f$ respectively. The highest points (or mid points of the corresponding intervals) of the two geodesics passing through the end points have angular coordinates $\xt_i, \xt_f$. We note that in general (unless the bulk curve has constant radial coordinate), the point at which the geodesic is tangent to bulk is not always its highest point and therefore $\xt_{i,f} \ne \tilde \xt_{i,f}$. The difference between the length of a bulk curve (with end points) and the (partial) differential entropy is compensated by the boundary terms that are associated with the (partial) differential entropy of a point. More precisely, the integral of differential entropy needs to be continued from $\xt_{f,i}$ to $\tilde \xt_{f,i}$ to remove the boundary terms. It is not difficult to understand this physical meaning of the boundary terms from the general definition of differential entropy. When the integral has end points, eq.\eq{Diffbound}, which is essentially the difference between the length of bulk curve and differential entropy reduces to the boundary term $S$, which by definition is the length of the geodesic starting from the tangent point on the bulk curve and ending on the boundary. 

People often use the other definition of differential entropy (as it is the original form proposed)
\be
\el{diff3d}
\tilde E[\xa(\xt)] := \frac 1 2 \oint\frac {d S}{d \xa} \rmd \xt\,,
\ee
the integrand of which is different from that of \eq{GdiffE} by a total derivative since the latter is given
by 
\be
E = \frac 1 2 \int\frac {d S}{d \xa} \rmd [\xt+\xa(\xt)] = \tilde E + \frac 1 2 \int d S\,.
\ee
The difference in the differential is due to $\xl_L = \xt + \xa(\xt)$ if
we choose $\xl = \xt$. The total derivative gives an extra boundary term given by the half length of the geodesic that is tangent to the end points. Due to this difference between $E$ and $\tilde E$, the physical meaning of $f(\xt, \tilde\theta)$ changes and now \cite{Czech:2014ppa} it represents the length of a segment on the geodesic from the end point (with angular coordinate $\tilde \xt$) to the mid point $\xt$. Note however that this discrepancy in the definition of differential entropy does not affects the validity of \eq{BHdiff}. 

\section{Construction of a co-dimension three object in higher dimensions}
Here we would like to generalize the construction in \cite{Czech:2014ppa} to higher dimensions in a very direct way. First of all, we note that the intervals for a point in AdS$_3$ are those spanned by all the geodesics that go through the point. In higher dimensions, the intervals for a hole become the minimal surfaces that are tangent to the bulk surface at a co-dimension three surface. A simple analogy then implies that a co-dimension three surface $P$ corresponds to all the minimal surfaces passing through it. As we shall see, this is indeed the case. Moreover, the infinitesimal distance between two of such objects $P_{A,B}$ can be computed from their respective differential entropies.

\subsection{General construction}
\label{sec:higherD}
Let us first look at the ``differential entropy" of a co-dimension three object
$P$ (which we will call extended point) in $d+1$ bulk space. As we see from \eq{Diffbound}, it can be expressed in terms of action angle variable
\begin{equation}
E(P)= \oint D\lambda\, \xG_\mu'p^\mu\at{s_b}\,.
\end{equation}
In this case, all the minimal surfaces go through $P$ (\ie $\xG(\xl,s_b) = P$) and therefore $\xG'(\xl, s_b) = 0$. This agrees with our intuitive understanding that the differential entropy for a ``point" is zero. Notice that $\xl$ here is no longer a number as the tangent vector in general can vary along $P$ and we need $d-1$ functions $\xl^\mu (\xs^a)$ ($a = 1, \dots d-2$) to parameterize the minimal surface passing $P$. Strictly speaking, $E(P)$ is not the differential
entropy in higher dimensions, which is by definition an integral of a single
variable
$\xl$ (instead of a function). However, as we shall see this new definition
correctly reproduce the infinitesimal distance between two objects $P_{A,B}$
and in the special case with translational symmetry, $E(P)$ reduces to the
differential entropy.

Now we move on to the computation of distance between two extended points. First of
all we recall that \cite{Czech:2014ppa} in AdS$_3$ the distance is related to the differential entropy
of $\xa_{AB} = \zt{min}\{\xa_A,\xa_B\}$ (which also follows from \eq{pointcurvedis}). In the language of integral geometry, we need to glue together two point curves $p_A$ and $p_B$ at $\xl_0$ ($\xa_A (\xl_0) = \xa_B (\xl_0)$) and the resulting differential entropy reproduces the distance. We are expecting a similar recipe in higher dimensions. The point is to find the transition point $\xl_0$ where the discontinuity occurs and accounts for distance since the differential entropy for a point
is a total derivative. A natural guess is where the difference in the integrand becomes zero
\be
\xD_{AB} \left.\frac{\xd S(\gamma_L(\lambda_0),\gamma_R(\lambda'))}{\xd
\lambda'}\right|_{\lambda'=\lambda_0} = 0\,.
\ee
Restricting to infinitesimal distance, we can replace $\xD_{AB}$ by derivative
and exchange the order of derivatives. As a result the transition point is at where
\be
v^\mu \frac {\xd} {\xd \xl} \frac {\p}{\p s_B^\mu} S \at{\xl_0}= v^\mu \frac {\xd p_\mu(s_B)} {\xd \xl} \at{\xl_0}= 0\,.
\ee
Such a discontinuity leads to a nonzero differential entropy
\be
\el{infdis}
E(P) = \oint D\xl \frac {\xd S} {\xd \xl} = \xD_{AB} S\at{\xl_0} = d_v
S= \int_P v^\mu p_\mu
\at{\xl_0}\,.
\ee
As we explained above, both $v^\mu$ and $p_\mu$ are functions along $P$ (parameterized
by $\xs$). The momentum $p_\mu$ by definition is a normalized vector 
\be
p_\mu = \frac {g_{\mu\nu} \dot \xg^\nu} {|\dot \xg|}\,,
\ee
and therefore is orthogonal to its derivative. As a result, at the transition
point $\xl_0$ we have $v^\mu$ parallel to $p^\mu$. In this case, $E(P)$ becomes
the gradient of the area functional along its tangent
vector and corresponds to the area of the infinitesimal co-dimension two bulk surface stretching between the bulk lines $P_A$ and $P_B$. In other words, we prove that the construction of $P$ does give the infinitesimal distance. Generalization to finite distance is straightforward as the differential entropy follows from the surface term $\xD_{AB} S|_{\xl_0}$ where
$\xl_0$ is the minimal surface connecting the two points $P_A,P_B$. The difference then gives the area of the minimal surface
between the two points.  

We would like to emphasize that the contribution to $E(P)$ follows from the
discontinuity at a particular configuration $\xl_0$. As a result, it is not
always necessary to do the integration over all possible $\xl$. For example,
when both $v$ and $P$ are translationally invariant, we only need to consider
minimal surfaces that are translationally invariant (in the $d-3$ longitudinal dimensions) and $\xl$ reduces to a variable (because it is independent of $\xs^a$). In this case $E(P)$ can be understood as differential entropy.    

\subsection{Example: a space with translational symmetry}
\label{subsec:HDtrans}
The discussion in Appendix~\ref{sec:higherD} may be a bit technical. So here we present a specific example with translational symmetry. In this case, $E(p)$ becomes the differential entropy and we can compute the integral explicitly. As we shall see, the result agrees with the general formula \eq{infdis} and correctly reproduces the bulk metric. It has been shown that the differential entropy for a bulk (co-dimension two) surface with translational symmetry can be expressed in a closed form \cite{Myers:2014jia}. So we will restrict ourselves to this case and study the bulk dual of a special set of strips,
which correspond to a co-dimension three surface in the bulk. This bulk object is a hyper-plane at $(x^1 = \tilde x, \tilde z)$, which is extended in the $x^i$-direction ($i=2,\dots d-1)$). 

Let us consider the general translationally invariant space \cite{Myers:2014jia},
\be
\rmd s^2=-g_0(z)\rmd t^2+\sum_{i=1}^{d-1}g_i(z)(\rmd x^i)^2+g_1(z)f(z)\rmd z^2\,.
\label{general}
\ee
Without lost of generality, we can pick $x^1$ as the transverse
dimension of the strip, which is translationally invariant in the other $d-2$
directions. We denote a strip by its center $x^1 = x$ and height $z_*$. We will see that the bulk geometry can still be reproduced from the variation of the ``differential entropy" associated with an extended point. First of all we note that the entanglement entropy still satisfies $4G \p_{\xD x} S =\sqrt {g(z_*)} := y(z_*)$ ($g:= \prod_{i=1}^{d-1} g_i$).

A direct application of the formula \eq{metricPoin} leads to
\ba
\label{linedz}
2G \int_{-\infty}^{+\infty} \frac {d^2 S} {d\xD x^2} \left|\frac {d\xD
x} {d\tilde z} \right|\rmd x \rmd z 
& = & -\int_{y(\tilde z)}^{0} \frac{\p x}{\p \tilde z}\at{y}\rmd y \rmd\tilde z =\rmd z\int_{0}^{y(\tilde z)}\frac{y\rmd y\sqrt{f(\tilde z)}}{\sqrt{g(\tilde z)-y^2}}\nn
& = & \sqrt{g(\tilde z)f(\tilde z)} \rmd z\,,
\ea
which precisely agrees with the line element along radial direction.
In the second equality we use 
\[\frac{\p y}{\p \tilde z}\at{x} = -\frac {\p x}{\p \tilde z}\at{y}/\frac {\p x}{\p y}\at{\tilde z}\,,\]
and the third equality follows from
\be
\frac {\p x}{\p \tilde z}\at{z_*}=\sqrt{g(z_*)}\frac{\sqrt{f(\tilde z)}}{\sqrt{g(\tilde z)-g(z_*)}}\,,
\ee
where $x$ is related to $\tilde z$ and $z_*$ by
\be
x-\tilde x =  \sqrt{g(z_*)}\int_{\tilde z}^{z_*}\rmd h\frac{\sqrt{f(h)}}{\sqrt{g(h)-g(z_*)}}\,.
\ee
So we obtain the ``distance" between two extended points, which is the area of the infinitesimal
$(d-1)$-surface swept out by the ``point" when it moves from $\tilde z$ to $\tilde z+d\tilde z$. As in the
$d=2$ case, such area is equal to the amount of entanglement entropy removed in the RG flow.


\begin{thebibliography}{99}
\bibitem{Ryu:2006bv} 
  S.~Ryu and T.~Takayanagi,
  ``Holographic derivation of entanglement entropy from AdS/CFT,''
  Phys.\ Rev.\ Lett.\  {\bf 96}, 181602 (2006)
  [hep-th/0603001].
    
\bibitem{Ryu:2006ef} 
  S.~Ryu and T.~Takayanagi,
  ``Aspects of Holographic Entanglement Entropy,''
  JHEP {\bf 0608}, 045 (2006)
  [hep-th/0605073].

\bibitem{VanRaamsdonk:2009ar} 
  M.~Van Raamsdonk,
  ``Comments on quantum gravity and entanglement,''
  arXiv:0907.2939 [hep-th].

\bibitem{VanRaamsdonk:2010}
  M.~Van Raamsdonk,
  ``Building up spacetime with quantum entanglement,''
  Gen.\ Rel.\ Grav.\  {\bf 42}, 2323 (2010)
  [Int.\ J.\ Mod.\ Phys.\ D {\bf 19}, 2429 (2010)]
  [arXiv:1005.3035 [hep-th]].

\bibitem{bianchimyers}
E.~Bianchi and R.~C.~Myers,
  ``On the architecture of spacetime geometry,''
  arXiv:1212.5183 [hep-th].

\bibitem{Maldacena:2013xja} 
  J.~Maldacena and L.~Susskind,
  ``Cool horizons for entangled black holes,''
  Fortsch.\ Phys.\  {\bf 61}, 781 (2013)
  [arXiv:1306.0533 [hep-th]].
  
\bibitem{MERA1}  
G. Vidal, ``A class of quantum many-body states that can be efficiently simulated",   Phys.\ Rev.\ Lett.\ {\bf 101}, 110501 (2008). [arXiv:quant-ph/0610099]

\bibitem{MERA2}
G. Evenbly, G. Vidal, ``Algorithms for entanglement renormalization", Phys.\ Rev.\ B {\bf 79}, 144108 (2009).   [arXiv:0707.1454 [cond-mat.str-el]].

\bibitem{Haegeman:2011uy} 
  J.~Haegeman, T.~J.~Osborne, H.~Verschelde and F.~Verstraete,
  ``Entanglement Renormalization for Quantum Fields in Real Space,''
  Phys.\ Rev.\ Lett.\  {\bf 110}, no. 10, 100402 (2013)
  [arXiv:1102.5524 [hep-th]].

\bibitem{Swingle:2009bg} 
  B.~Swingle,
  ``Entanglement Renormalization and Holography,''
  Phys.\ Rev.\ D {\bf 86}, 065007 (2012)
  [arXiv:0905.1317 [cond-mat.str-el]].

\bibitem{Evenbly}
G. Evenbly, G. Vidal, ``Tensor network states and geometry", J.\ Stat.\ Phys.\ (2011) 145:891-918, arXiv:1106.1082 [quant-ph].

\bibitem{Swingle:2012} 
  B.~Swingle,
  ``Constructing holographic spacetimes using entanglement renormalization,''
  arXiv:1209.3304 [hep-th].
  
\bibitem{Nozaki:2012zj} 
  M.~Nozaki, S.~Ryu and T.~Takayanagi,
  ``Holographic Geometry of Entanglement Renormalization in Quantum Field Theories,''
  JHEP {\bf 1210}, 193 (2012)
  [arXiv:1208.3469 [hep-th]].

\bibitem{Qi:2013caa} 
  X.~L.~Qi,
  ``Exact holographic mapping and emergent space-time geometry,''
  arXiv:1309.6282 [hep-th].
  
\bibitem{Lee:2015vla} 
  C.~H.~Lee and X.~L.~Qi,
  ``Exact holographic mapping in free fermion systems,''
  arXiv:1503.08592 [hep-th].


\bibitem{Mollabashi:2013lya} 
  A.~Mollabashi, M.~Nozaki, S.~Ryu and T.~Takayanagi,
  ``Holographic Geometry of cMERA for Quantum Quenches and Finite Temperature,''
  JHEP {\bf 1403}, 098 (2014)
  [arXiv:1311.6095 [hep-th]].

\bibitem{Pastawski:2015} 
  F.~Pastawski, B.~Yoshida, D.~Harlow and J.~Preskill,
  ``Holographic quantum error-correcting codes: Toy models for the bulk/boundary correspondence,''
  arXiv:1503.06237 [hep-th].

\bibitem{Bao:2015uaa} 
  N.~Bao, C.~Cao, S.~M.~Carroll, A.~Chatwin-Davies, N.~Hunter-Jones, J.~Pollack and G.~N.~Remmen,
  ``Consistency Conditions for an AdS/MERA Correspondence,''
  arXiv:1504.06632 [hep-th].

\bibitem{Miyaji:2014mca} 
  M.~Miyaji, S.~Ryu, T.~Takayanagi and X.~Wen,
  ``Boundary States as Holographic Duals of Trivial Spacetimes,''
  JHEP {\bf 1505}, 152 (2015)
  [arXiv:1412.6226 [hep-th]].

\bibitem{Miyaji:2015yva} 
  M.~Miyaji and T.~Takayanagi,
  ``Surface/State Correspondence as a Generalized Holography,''
  arXiv:1503.03542 [hep-th].
    
\bibitem{Verlinde:2015qfa} 
  H.~Verlinde,
  ``Poking Holes in AdS/CFT: Bulk Fields from Boundary States,''
  arXiv:1505.05069 [hep-th].
  
\bibitem{Miyaji:2015fia} 
  M.~Miyaji, T.~Numasawa, N.~Shiba, T.~Takayanagi and K.~Watanabe,
  ``cMERA as Surface/State Correspondence in AdS/CFT,''
  arXiv:1506.01353 [hep-th].


\bibitem{ooguri} 
  Y.~Nakayama and H.~Ooguri,
  ``Bulk Locality and Boundary Creating Operators,''
  arXiv:1507.04130 [hep-th].
  
  
\bibitem{Hamilton:2006} 
  A.~Hamilton, D.~N.~Kabat, G.~Lifschytz and D.~A.~Lowe,
  ``Holographic representation of local bulk operators,''
  Phys.\ Rev.\ D {\bf 74}, 066009 (2006)
  [hep-th/0606141].

\bibitem{bilson1} 
  S.~Bilson,
  ``Extracting spacetimes using the AdS/CFT conjecture,''
  JHEP {\bf 0808}, 073 (2008)
  [arXiv:0807.3695 [hep-th]].

\bibitem{bilson2} 
  S.~Bilson,
  ``Extracting Spacetimes using the AdS/CFT Conjecture: Part II,''
  JHEP {\bf 1102}, 050 (2011)
  [arXiv:1012.1812 [hep-th]].
  
\bibitem{Spillane} 
  M.~Spillane,
  ``Constructing space from entanglement entropy,''
  arXiv:1311.4516 [hep-th].

\bibitem{Bousso:2012} 
  R.~Bousso, S.~Leichenauer and V.~Rosenhaus,
  ``Light-sheets and AdS/CFT,''
  Phys.\ Rev.\ D {\bf 86}, 046009 (2012)
  [arXiv:1203.6619 [hep-th]].

\bibitem{Czech:2012}
  B.~Czech, J.~L.~Karczmarek, F.~Nogueira and M.~Van Raamsdonk,
  ``The gravity dual of a density matrix,''
  Class.\ Quant.\ Grav.\  {\bf 29}, 155009 (2012)
  [arXiv:1204.1330 [hep-th]].

\bibitem{Chen:2014ec} 
Y.~Chen and G.~Vidal, 
``Entanglement contour," 
J. Stat. Mech. (2014) P10011.
  [arXiv:1406.1471 [cond-mat.str-el]].

\bibitem{Balasubramanian:2013b}
  V.~Balasubramanian, B.~D.~Chowdhury, B.~Czech and J.~de Boer,
  ``The entropy of a hole in spacetime,''
  JHEP {\bf 1310}, 220 (2013)
  [arXiv:1305.0856 [hep-th]].

\bibitem{Balasubramanian:2013lsa} 
  V.~Balasubramanian, B.~D.~Chowdhury, B.~Czech, J.~de Boer and M.~P.~Heller,
  ``Bulk curves from boundary data in holography,''
  Phys.\ Rev.\ D {\bf 89}, no. 8, 086004 (2014)
  [arXiv:1310.4204 [hep-th]].

\bibitem{Myers:2014jia} 
  R.~C.~Myers, J.~Rao and S.~Sugishita,
  ``Holographic Holes in Higher Dimensions,''
  JHEP {\bf 1406}, 044 (2014)
  [arXiv:1403.3416 [hep-th]].

\bibitem{Czech:2014wka} 
  B.~Czech, X.~Dong and J.~Sully,
  ``Holographic Reconstruction of General Bulk Surfaces,''
  JHEP {\bf 1411}, 015 (2014)
  [arXiv:1406.4889 [hep-th]].

\bibitem{Headrick:2014eia} 
  M.~Headrick, R.~C.~Myers and J.~Wien,
  ``Holographic Holes and Differential Entropy,''
  JHEP {\bf 1410}, 149 (2014)
  [arXiv:1408.4770 [hep-th]].

\bibitem{Czech:2014ppa} 
  B.~Czech and L.~Lamprou,
  ``Holographic definition of points and distances,''
  Phys.\ Rev.\ D {\bf 90}, no. 10, 106005 (2014)
  [arXiv:1409.4473 [hep-th]].

\bibitem{santalo}
  L.~Santal{\'o},
  ``Integral geometry and geometric probability,''
  Cambridge University Press, 1976.

\bibitem{Czech:2015qta} 
  B.~Czech, L.~Lamprou, S.~McCandlish and J.~Sully,
  ``Integral Geometry and Holography,''
  arXiv:1505.05515 [hep-th].

\bibitem{Lin:2014hva} 
  J.~Lin, M.~Marcolli, H.~Ooguri and B.~Stoica,
  ``Locality of Gravitational Systems from Entanglement of Conformal Field Theories,''
  Phys.\ Rev.\ Lett.\  {\bf 114}, no. 22, 221601 (2015)
  [arXiv:1412.1879 [hep-th]].
  
\bibitem{Nozaki:2013wia} 
  M.~Nozaki, T.~Numasawa and T.~Takayanagi,
  ``Holographic Local Quenches and Entanglement Density,''
  JHEP {\bf 1305}, 080 (2013)
  [arXiv:1302.5703 [hep-th]].

\bibitem{Bhattacharya:2014vja} 
  J.~Bhattacharya, V.~E.~Hubeny, M.~Rangamani and T.~Takayanagi,
  ``Entanglement density and gravitational thermodynamics,''
  Phys.\ Rev.\ D {\bf 91}, no. 10, 106009 (2015)
  [arXiv:1412.5472 [hep-th]].

\bibitem{Czech:2014tva} 
  B.~Czech, P.~Hayden, N.~Lashkari and B.~Swingle,
  ``The Information Theoretic Interpretation of the Length of a Curve,''
  arXiv:1410.1540 [hep-th].

\bibitem{Jackson:2013eqa} 
  S.~Jackson, R.~Pourhasan and H.~Verlinde,
  ``Geometric RG Flow,''
  arXiv:1312.6914 [hep-th].
  
\bibitem{Santalo:1952}
Santal\'o, 
"Measure of sets of geodesics in a Riemannian space and applications to integral formulas in elliptic and hyperbolic spaces," Summa Brasil. Math. {\bf 13} (1952), fasc. 1.

\bibitem{Porrati:2003na} 
  M.~Porrati and R.~Rabadan,
  ``Boundary rigidity and holography,''
  JHEP {\bf 0401}, 034 (2004)
  [hep-th/0312039].

\bibitem{pestov:2005}
L.~Pestov and G.~Uhlmann,
``Two-dimensional compact simple Riemannian manifolds are boundary distance rigid,''
Ann. Math. {\bf 161} 2 (2005), 1093-1110.

\bibitem{Hayden:2011ag} 
  P.~Hayden, M.~Headrick and A.~Maloney,
  ``Holographic Mutual Information is Monogamous,''
  Phys.\ Rev.\ D {\bf 87}, no. 4, 046003 (2013)
  [arXiv:1107.2940 [hep-th]].

\end{thebibliography}
\end{document}